\documentclass[10pt,twocolumn]{article}
\usepackage[top=1.9cm, bottom=1.9cm, left=1.8cm, right=1.8cm]{geometry}

\usepackage{amsmath,amssymb,amsfonts}
\usepackage{listings}                       %
\usepackage[noend]{algpseudocode}
\usepackage{multirow}
\usepackage{graphicx}
\usepackage{textcomp}
\usepackage{xcolor}
\usepackage{flushend}
\usepackage[lined,boxed]{algorithm2e}

\usepackage{graphicx,subfigure}
\usepackage[small,bf]{caption}
\usepackage[square,sort,comma,numbers]{natbib}

\SetCommentSty{mycommfont}

\usepackage{xcolor}
\usepackage{booktabs}
\usepackage{graphicx}
\usepackage{paralist}

\usepackage{times}
\renewcommand{\footnotesize}{\fontsize{8}{9}\selectfont}
\usepackage[compact]{titlesec}
\titlespacing*{\section}{0pt}{*4}{4pt}
\titlespacing{\subsection}{0pt}{*3}{3pt}
\titlespacing{\subsubsection}{0pt}{*3}{3pt}

\definecolor{linkcol}{rgb}{0,0,0.5}
\definecolor{citecol}{rgb}{0,0.5,0.3}
\definecolor{urlcol}{rgb}{0.3,0,0}

\usepackage{xspace}

\renewenvironment{thebibliography}[1]{
  \begin{oldthebibliography}{#1}
    \setlength{\itemsep}{0.0em}
    \setlength{\parskip}{0.0em}
}
{
  \end{oldthebibliography}
}

\usepackage[hang,flushmargin]{footmisc}

\renewcommand{\footnoterule}{%
  \kern -3pt
  \hrule width 1in
  \kern 2pt
}

\usepackage{url}
\makeatletter
\def\url@leostyle{%
  \@ifundefined{selectfont}{\def\UrlFont{}}%
  {\def\UrlFont{}}%
}
\makeatother
\usepackage[hyphenbreaks]{breakurl}

\definecolor{darkred}{RGB}{153,0,0}
\definecolor{darkblue}{RGB}{0,0,99}
\usepackage[colorlinks=true, linkcolor = darkred,   citecolor = darkred, urlcolor = darkblue]{hyperref}
\newcommand{\edit}[1]{\textcolor{black}{#1}}

\newtheorem{ldp-definition}{Definition}
\newtheorem{dp-definition}{Definition}
\newcommand{\reduce}{\vspace*{-0.00cm}}
\newcommand{\descr}[1]{\smallskip\noindent\textbf{#1}}
\newcommand{\descrit}[1]{\vspace{0.05cm}\noindent\textbf{\em #1}}

\newcommand{\ndss}[1]{\textcolor{black}{#1}}

\begin{document}

\sloppy 

\title{\bf Local and Central Differential Privacy for Robustness\\ and Privacy in Federated Learning\thanks{Published in the Proceedings of the 29th Network and Distributed System Security Symposium (NDSS 2022). Please cite accordingly.
}}

\author{Mohammad Naseri$^1$, Jamie Hayes$^{2}$, and Emiliano De Cristofaro$^{1,3}$\\[0.5ex]
\fontsize{11}{12}\selectfont $^1$University College London\;\; $^2$DeepMind\;\; $^3$Alan Turing Institute}
\date{}

\maketitle

\begin{abstract}
Federated Learning (FL) allows multiple participants to train machine learning models collaboratively by keeping their datasets local while only exchanging model updates.
Alas, this is not necessarily free from privacy and robustness vulnerabilities, e.g., via membership, property, and backdoor attacks. 
This paper investigates whether and to what extent one can use differential Privacy (DP) to protect both privacy and robustness in FL.
To this end, we present a first-of-its-kind evaluation of Local and Central Differential Privacy (LDP/CDP) techniques in FL, assessing their feasibility and effectiveness.

Our experiments show that both DP variants do defend against backdoor attacks, albeit with varying levels of protection-utility trade-offs, but anyway more effectively than other robustness defenses.
DP also mitigates {\em white-box} membership inference attacks in FL, and our work is the first to show it empirically.
Neither LDP nor CDP, however, defend against property inference.
Overall, our work provides a comprehensive, re-usable measurement methodology to quantify the trade-offs between robustness/privacy and utility in differentially private FL.
\end{abstract}

\section{Introduction}
Aiming to increase privacy and communication efficiency, {\em Federated Learning} (FL)~\cite{mcmahan2017communication} has emerged as a compromise between a centralized approach to Machine Learning (ML), where training data is pooled at a central server, and the alternative of only training local models, client-side. 
Multiple ``participants'' collaborate in solving an ML problem by keeping their data on their device and only exchanging model parameters~\cite{kairouz2019advances}. %
Currently, FL is used in many different settings, e.g., Google's predictive keyboard~\cite{hard2018federated, yang2018applied}, voice assistants~\cite{bhowmick2018protection},  emoji prediction~\cite{ramaswamy2019federated}, healthcare~\cite{sheller2018multi,brisimi2018federated}, etc.

Alas, previous work has highlighted robustness and privacy weaknesses of FL~\cite{kairouz2019advances}.
For instance, poisoning attacks may reduce the model's accuracy or make it misbehave on specific inputs.
In particular, in {\em backdoor} attacks, a malicious client injects a backdoor into the final model to corrupt the performance of the trained model on specific sub-tasks~\cite{bagdasaryan2020backdoor}.
Moreover, an adversary may be able to infer {\em membership}~\cite{nasr2019comprehensive} (i.e., learn if a data point is part of a target's training set), or {\em properties}~\cite{melis2019exploiting} of the training data.

Prior work has investigated robustness defenses primarily based on robust, Byzantine-robust aggregation algorithms~\cite{blanchard2017machine,yin2018byzantine,shejwalkar2021manipulating}.
As for privacy, FL techniques have since their inception supported homomorphic encryption so that the server can only decrypt the aggregates~\cite{bonawitz2017practical}, but this does not eliminate leakage from the aggregates~\cite{melis2019exploiting}.
The established framework to define functions that are free from adversarial inferences is Differential Privacy (DP)~\cite{dwork2014algorithmic}, 
allowing to bound the privacy loss of individual data subjects by adding noise.
In the context of FL, one can use two DP variants:
1) Local DP (LDP)~\cite{pihur2018differentially}, where each participant adds noise before sending updates to the server;
and 2) Central DP (CDP)~\cite{mcmahan2017learning, geyer2017differentially}, where %
it is the server to apply a DP aggregation algorithm.

\descr{Problem Statement.}
Backdoor attacks in FL are first explored by Bagdasaryan et al.~\cite{bagdasaryan2020backdoor}. 
However,~\cite{bagdasaryan2020backdoor} does not present any working defense; instead, they discuss how existing ones (including participant-level CDP) are not suited to FL.
Sun et al.~\cite{sun2019can} introduce two defenses against backdoor attacks in FL: \ndss{bounding the norm of gradient updates and adding Gaussian noise, aiming to reduce the effect of poisonous data.
Although they are not meant to provide privacy, we believe they could, in theory, be used for this purpose because of the noisy updates; however, we show that, in practice, they do not defend against inference attacks.}

\ndss{Consequently, our work sets out to tackle the following research question: can we deploy defenses to mitigate {\em both} backdoor {\em and} inference attacks? If so, how and with what utility trade-offs?
We investigate whether Differential Privacy, both in its CDP and LDP instantiations, can be used for this purpose, considering different scenarios of how they can be applied and providing an extensive measurement study of their real-world effectiveness.}
Our main intuition is that CDP limits the information exposed about a specific participant, while LDP does so for records in a participant's dataset; in both cases, the impact of poisonous data should be reduced while simultaneously protecting against inference attacks.

\descr{Technical Roadmap.}  We introduce an analytical approach to understand the effectiveness of LDP and CDP to protect federated models from both backdoor and inference attacks while maintaining good utility.
This entails addressing a few challenges. 
First, we do not know how to compare the protection they yield, as they are not meant to guarantee robustness, and even for privacy, their definitions capture slightly different notions.
Moreover, there is no straightforward analytical way to determine the effect of LDP/CDP on utility. %

We run experiments on three datasets for backdoor attacks: EMNIST (handwritten digits), CIFAR10 (different classes of images), Reddit comments, and \ndss{Sentiment140 (tweets)}. %
We consider two FL settings with varying numbers of participants/attackers.
Further, we experiment with white-box\footnote{White-box refers to the attacker having complete knowledge of the system's parameters, except for the participants' datasets.} membership inference attacks~\cite{nasr2019comprehensive} on the CIFAR100 dataset (different classes of images), Purchase100 (records of purchases), and  \ndss{Texas100 (records of hospital discharges).}
We consider both active and passive attacks, run from both server and participant sides. 
Finally, we run property inference attacks~\cite{melis2019exploiting} for a gender classification task on the LFW dataset.

\descr{Main Findings.} In summary, we find that:
\begin{compactitem} 
\item Both LDP and CDP do defend against backdoor attacks, albeit with varying levels of protection and utility, but overall better than prior work.
\ndss{\item Applying LDP only on non-attackers in FL can actually boost the backdoor attack accuracy.}
\item Both LDP and CDP also protect against (white-box) membership inference~\cite{nasr2019comprehensive}; although this is not entirely unexpected, we are the first to show that this does not necessarily come with a high cost on utility.
\item LDP does not work against property inference attacks. %
Although CDP can, in theory, defend against the attack, it does so with a significant loss in utility.
\end{compactitem} 

\noindent More precisely, our robustness experiments show that, with 2,400 participants on EMNIST, and with 1\% of them selected at every round and 1 attacker performing the backdoor attack,  LDP and CDP (with $\epsilon=3$) reduce the accuracy of the attack from 88\% to 10\% and 6\%, respectively, with only a limited reduction in utility.
By comparison, using defenses from~\cite{sun2019can} (see Section~\ref{sec:poisoning-defenses}), attack accuracy only goes down to 37\% with norm bounding and 16\% with so-called weak DP, which does not provide privacy, unlike LDP/CDP.

Our membership inference experiments show that, with 4 participants (the same setting as~\cite{nasr2019comprehensive}), LDP ($\epsilon=8.6$) and CDP ($\epsilon=5.8$) reduce the accuracy of an active (local) attack from 75\% to 55\% and 52\%, respectively, on CIFAR100, and from 68\% to 54\% and 55\% on Purchase100 (50\% baseline).
As mentioned, LDP is ineffective against property inference, especially when the target participant has many data points with the property; this is because LDP only provides record-level privacy.
Although CDP can, in theory, defend against the attack, it does so with a significant loss in utility.
For instance, with 10 participants on LFW,  CDP only yields a 57\% accuracy on the main task when $\epsilon=4.7$; with $\epsilon=8.1$, main task accuracy reaches 83\%, but with no prediction, the accuracy of the property inference only goes down from 87\% to 85\%.\footnote{Unlike~\cite{melis2019exploiting}, we are able to make our models converge by increasing the privacy budget (e.g., $\epsilon > 8$), %
although this is not enough to thwart the attack.}

\descr{Contributions.} The main contributions of our work include:
\begin{compactenum}
\item \ndss{We are the first to propose the use of LDP to mitigate backdoor attacks against FL. 
In addition, we are the first to experimentally show that CDP mitigates backdoor attacks in FL.}
In fact, both LDP and CDP can defend better than the state of the art~\cite{sun2019can} (which used norm bounding and weak DP), additionally protecting privacy as well, while only slightly decreasing utility.\smallskip
\item We are the first to show that both LDP and CDP can effectively (i.e., without destroying utility) defend against white-box membership inference attacks in FL~\cite{nasr2019comprehensive}, and more so than theoretically expected.\smallskip

\item We provide a re-usable measurement framework to quantify the trade-offs between robustness/privacy and utility yielded by LDP and CDP in FL.

\end{compactenum}

\reduce\reduce\reduce
\section{Preliminaries}
\reduce
We now review Federated Learning (FL) and Differential Privacy (DP).
Readers who are already familiar with these concepts can skip this section without loss of continuity.

\reduce\reduce
\subsection{Federated Learning (FL)}
\reduce
FL is a collaborative learning setting to train machine learning models~\cite{mcmahan2017communication}.
It involves $N$ participants (or clients), each holding their own private dataset, and a central server (or aggregator).
Unlike the traditional centralized approach, data is not pooled at a central server; whereas, participants train models locally and exchange updated parameters with the server, which aggregates and sends them to the participants.

FL involves multiple rounds:
In round 0, the server generates a model with random parameters $\theta_0$, which is sent to all participants.
Then, at each round $r$, $K$ out of $N$ participants are selected at random; each participant $i$ locally computes training gradients according to their local dataset $D_i$ and sends the updated parameters to the server. 
The latter computes the global parameters $\theta_r=\Sigma_{i=1}^{K}\theta_i/K$ and sends them to all $N$ participants for the next round.
After a certain number of rounds ($R$), the model is finalized with parameters $\theta_R$.

There are different privacy-enhancing techniques used in FL. 
Homomorphic Encryption (HE) can be used to encrypt participants' parameters in such a way that the server can only decrypt the aggregates~\cite{bonawitz2017practical, cheng2019secureboost, hardy2017private}. %
However, this is resource-intensive and does not mitigate inference attacks over the output of the aggregation, as global parameters can still leak information~\cite{kairouz2019advances}.
Another approach is to use differentially private techniques, which we review next.

\reduce\reduce
\subsection{Differential Privacy (DP)}\label{sec:DP}
\reduce
Differential Privacy provides statistical guarantees against the information an adversary can infer through the output of a randomized algorithm.
It provides an unconditional upper bound on the influence of a single individual on the output of the algorithm by adding noise~\cite{dwork2014algorithmic}. 

\begin{dp-definition}
{\bf Differential Privacy.} A randomized mechanism $M$ provides $(\epsilon, \delta)$-{\em differential privacy} if for any two neighboring databases, $D_1$ and $D_2$, that differ in only a single record, and for all possible outputs $S \subseteq Range(A)$: \vspace{-0.4cm}

   \begin{equation}
     	\vspace{-0.4cm}
	P[M(D_1 \in A)]\leq e^{\epsilon} 	P[M(D_2 \in A)] + \delta %
   \end{equation}
 \end{dp-definition}
 
The $\epsilon$ parameter (aka {\em privacy budget}) is a metric of privacy loss.
It also controls the privacy-utility trade-off, i.e., lower $\epsilon$ values indicate higher levels of privacy, but likely reduce utility too.
The $\delta$ parameter accounts for a (small) probability on which the upper bound $\epsilon$ does not hold.
The amount of noise needed to achieve DP is proportional to the {\em sensitivity} of the output; this measures the maximum change in the output due to the inclusion or removal of a single record.

In ML, DP is used to learn a distribution of a dataset while providing privacy for individual records~\cite{ji2014differential}.
Differentially Private Stochastic Gradient Descent (DP-SGD)~\cite{abadi2016deep}, and Private Aggregation of Teacher Ensembles (PATE)~\cite{papernot2016semi} are two different approaches to privacy-preserving ML; in this paper, we use the former.
DP-SGD~\cite{abadi2016deep} uses a noisy version of stochastic gradient descent to find differentially private minima for the optimization problem.
This is done by bounding the gradients and then adding noise with the help of the ``moments accountant'' technique to keep track of the spent privacy budget.
Whereas PATE~\cite{papernot2016semi} provides privacy for training data using a student-teacher architecture.

\begin{algorithm}[t]
\footnotesize
	\DontPrintSemicolon
  \SetKwFunction{FMain}{Main}	
   \SetKwFunction{FDPSGD}{DP-SGD}
  \SetKwProg{Fn}{Function}{:}{\Return}
  \Fn{\FMain{}}{
        Initialize: model $\theta_0$\;
        \For{each round $r=1,2,...$}{
	        $K_r\gets $ randomly select $K$ participants\;
	        \For{each participant $k \in K_r$}{
		        $\theta_{r} ^ {k} \gets$ DP-SGD(\ldots) \tcp*{This is done in parallel}
	        }
			$\theta_r \gets  \Sigma_{i=1}^{K_r} {\dfrac{n^k}{n} \theta_r^k}$\tcp*{$n^k$ is the size of $k$'s dataset} 
        }
        }
\;
  \SetKwProg{Pn}{Function}{:}{\Return$\theta_E$}
  \Pn{\FDPSGD{Clipping norm $S$, dataset $D$, sampling probability $p$,
	 noise magnitude $\sigma$, learning rate $\eta$, Iterations $E$, loss function $L(\theta(x), y)$}}{
        Initialize $\theta_0$\;
       	\For{each local epoch $i$ from 1 to E}{
       		\For{($x$, $y$) $\in$ random $batch$ from dataset $D$ with probability $p$}{
    				$g_i=\nabla_{\theta} L(\theta_i;(x,y))$\;
    			}
    			$Temp = \dfrac{1}{pD}{\Sigma_{i\in batch} g_i min(1, \dfrac{S}{\| g_i\|_2}) + N(0, \sigma^2I)}$ \;
    			$\theta_{i+1}=\theta_{i} - \eta(Temp)$ \;
    		}
    		
  }
\caption{Local Differential Privacy in FL.}\label{alg:ldpalgo}
\end{algorithm}
  	\vspace{-0.2cm}

\begin{algorithm}[ht]
\footnotesize
	\DontPrintSemicolon
  \SetKwFunction{FMain}{Main}
   \SetKwFunction{FPUPDATE}{Participant\_Update}
  \SetKwProg{Fn}{Function}{:}{\Return}
  \Fn{\FMain{}}{
		Initialize: model $\theta_0$, Moment\_Accountant($\epsilon$, N)\tcp*{N is the number of all participants}
        \For{each round $r=1,2,...$}{
				$C_r\gets $ randomly select participants with probability q\;
				$p_r\gets $Moment\_Accountant.get\_privacy\_spent() \tcp*{Returns the spent privacy budget for the current round}
				\If{$p_r > $ $T$\tcp*{If the spent privacy budget is greater than the threshold, return the current model}}{ 
					return $\theta_r$\;
				}
				\For{each participant $k \in C_r$} {
					$\Delta_k ^ {r+1} \gets$Participant\_Update({$k, \theta_r$})\tcp*{This is done in parallel}
		
				}
	$S\gets {bound}$\;
	$z\gets {noise\_scale}$\;
	$\sigma\gets {zS/q}$\;
	$\theta_{r+1}\gets {\theta_r + \Sigma_{i=1}^{C_r}{\Delta_i ^ {r+1} }/C_r +  N (0, I\sigma^2)}$\;
	Moment\_Accountant.accumulate\_spent\_privacy($z$)\;

        }
        }
  \;
  \SetKwProg{Pn}{Function}{:}{\Return$\theta-\theta_r$\tcp*{This one is already clipped\reduce}}
  \Pn{\FPUPDATE{$k$, $\theta_r$}}{
  		$\theta\gets {\theta_r}$\;
  		\For{each local epoch $i$ from 1 to E}{
    				\For{batch $b \in B$}{
    					$\theta \gets \theta - \eta\nabla L(w;b)$\;
					$\Delta \gets \theta - \theta_r$	\;		
					$\theta \gets \theta_0 + \Delta \min(1,\dfrac{S}{\| \Delta\|_2 })$\;
    				}
    		}
  }
	\caption{Central Differential Privacy in FL.}\label{alg:cdpalgo}
  	\vspace{-0.2cm}
\end{algorithm}

\reduce\reduce
\subsection{DP in FL}\label{sec:DPFL}
\reduce

As mentioned, in the context of FL, one can use one of two variants of DP, namely, {\em local} and {\em central}~\cite{mcmahan2017learning, geyer2017differentially,pihur2018differentially}.

\descr{Local Differential Privacy (LDP).}
With LDP, the noise addition required for DP is performed locally by each participant.
Each participant runs a random perturbation algorithm $M$ and sends the results to the server.
The perturbed result is guaranteed to protect an individual's data according to the $\epsilon$ value.
This is formally defined next~\cite{duchi2013local}.\reduce

\begin{dp-definition}
   Let X be a set of possible values and Y the set of noisy values. $M$ is $(\epsilon, \delta)$-{\em locally differentially private} ($\epsilon$-LDP) if for all $x_1, x_2$ $\in X$ and for all $y \in Y$:\vspace{-0.1cm}
   \begin{equation}
	P[M(x)=y]\leq e^{\epsilon} P[M(x')=y] + \delta\vspace{-0.1cm}
   \end{equation}
 \end{dp-definition}\vspace{-0.2cm}

\noindent We implement LDP in FL because participants use differentially private stochastic gradient descent (DP-SGD)~\cite{abadi2016deep} to train the model on their datasets.
\edit{This approach allows us to use moments accountant to keep track of the spent privacy budget.
}
Algorithm~\ref{alg:ldpalgo} shows how DP-SGD in LDP works.

\descr{Central Differential Privacy (CDP).}
With CDP, the FL aggregation function is perturbed by the server, and this provides {\em participant-level} DP.
This guarantees that the output of the aggregation function is indistinguishable, with probability bounded by $\epsilon$, to whether or not a given participant is part of the training process.
In this setting, participants need to trust the server: 1) with their model updates, and 2) to correctly perform perturbation by adding noise, etc. 
While some degree of trust in the server is needed, this is a much weaker assumption than entrusting the server with the data itself.
If anything, inferring training set membership or properties from the model updates is much less of a significant privacy threat than having data in the clear.
Moreover, in FL, clients do not share entire datasets also for efficiency reasons and/or because they might be unable to for policy or legal reasons.

In this paper, we implement the CDP approach for FL discussed in~\cite{mcmahan2017learning} and \cite{geyer2017differentially}, which is illustrated in Algorithm~\ref{alg:cdpalgo}.
The server clips the $l_2$ norm of participants' updates, 
then it aggregates the clipped updates and adds Gaussian noise to the aggregate.
This prevents overfitting to any participant's updates. 
To track the privacy budget spent, the moments accountant technique from in~\cite{abadi2016deep} can be used.

\reduce\reduce
\section{Defending Against Backdoor Attacks in FL}\label{Poisoning Attacks}
\reduce

In this section, we experiment with LDP and CDP against backdoor attacks, while also comparing our results with state-of-the-art defenses in terms of both robustness and utility. 

\reduce\reduce
\subsection{Backdoor Attack}\label{sec:pois-att}
\reduce

Backdoor attacks are a special kind of {\em poisoning} attacks. We first review the latter, then formally define the former.

\descr{Poisoning attacks} can be divided into random and targeted ones. In the former, the attacker aims to decrease the accuracy of the final model; in the latter, the goal is to make the model output a target label pre-defined by the attacker~\cite{huang2011adversarial}.
In the context of FL, participants (and not the server) are potential adversaries. 
Random attacks are easier to identify as the server could check if the accuracy is below a threshold or not.

These attacks can be performed on data or models. 
In the former, the attacker adds examples to the training set to modify the final model's behavior.
In the latter, she poisons the local model before sending it to the server.
In both cases, the goal is to make the final model misclassify a set of predefined inputs.
Remind that, in FL, data is never sent to the server. 
Therefore, anything that can be achieved with poisonous data is also feasible by poisoning the model~\cite{enthoven2020overview,lyu2020threats}.

\descr{Backdoor attacks.} A malicious client injects a backdoor task into the final model. 
Following~\cite{bagdasaryan2020backdoor, bhagoji2019analyzing, sun2019can}, we consider targeted model poisoning attacks and refer to them as backdoor attacks. 
These attacks in FL are relatively straightforward to implement and rather effective~\cite{bagdasaryan2020backdoor}; it is not easy to defend against them, as the server cannot access participants' data as that would violate one of the main principles of FL.

The main intuition is to rely on a model-replacement methodology, similar to~\cite{sun2019can,bagdasaryan2020backdoor}. %
In round $r$, the attacker attempts to introduce a backdoor and replaces the aggregated model with a backdoored one $\theta^*$, by sending the following model update to the server:\vspace{-0.1cm}
 \begin{equation}
	\Delta\theta_{r}^{attacker} = \dfrac{\Sigma_{i=1}^{K}{n_i}}{\eta n_{attacker}} \cdot (\theta^* - \theta_r)\vspace{-0.1cm}
   \end{equation}
where $n_i$ is \#data points at participant i and $\eta$ the server learning rate. 
Then, the aggregation in the next round yields:\vspace{-0.1cm}
\begin{equation}
	\Delta\theta_{r+1} = \theta^* + \eta\dfrac{\Sigma_{i=1}^{K-1}{n_i \Delta\theta_{r}^i}}{\Sigma_{i=1}^{K}n_i}\vspace{-0.1cm}
   \end{equation}
If we assume the training process is in its last rounds, then the aggregated model is going to converge; therefore, model updates from non-attacker participants are small, and we would have $\Delta\theta_r \simeq \theta^*$.

\reduce\reduce
\subsection{Defenses}\label{sec:poisoning-defenses}
\reduce

One straightforward defense against poisoning attacks is byzantine-resilient aggregation frameworks, e.g., Krum~\cite{blanchard2017machine} or coordinate-wise median~\cite{yin2018byzantine}. 
However, as showed in~\cite{bhagoji2019analyzing}, these are not effective in the FL setting.
Overall, FL is vulnerable to backdoor attacks as it is difficult to control the local model submitted by malicious participants. 
Defenses that require access to the training data are not applicable as that would violate the privacy of the participants, defeating one of the main reasons to use FL in the first place. 
In fact, even defenses that do not require training data, e.g., DeepInspect~\cite{chen2019deepinspect},  require inverting the model in order to extract the training data, thus violating the privacy-preserving goal of FL.
Similarly, if a model is trained over encrypted data, e.g., using CryptoNet~\cite{gilad2016cryptonets} or SecureML~\cite{mohassel2017secureml}, the server cannot detect anomalies in participant's updates.

Sun et al.~\cite{sun2019can} show that the performance of backdoor attacks in FL ultimately depends on the fraction of adversaries and the complexity of the task. 
They also propose two defenses: 1) {\em Norm Bounding} and 2) {\em Weak DP}.
\ndss{Byzantine-robust defenses like Krum~\cite{yin2018byzantine}, Trimmed Mean~\cite{yin2018byzantine}, or Divide-and-Conquer (DnC)~\cite{shejwalkar2021manipulating} are designed to defend against robustness attacks but not to provide privacy. 
Unlike them, norm bounding and ``Weak DP'' might potentially provide both robustness and privacy and are thus the main focus in this paper for comparisons.}

\descr{Norm Bounding.} 
If model updates received from attackers are over some threshold, then the server can simply ignore those participants.
However, if the attacker is aware of the threshold, it can return updates within that threshold.
Sun et al.~\cite{sun2019can} assume that the adversary has this strong advantage and apply norm bounding as a defense, guaranteeing that the norm of each model update is small.
If we assume that the updates' threshold is $T$, then the server can ensure that norms of participants' updates are within the threshold:\vspace{-0.1cm}
 \begin{equation}
	\Delta\theta_{r+1} {=} \sum_{i=1}^{k}{\dfrac{\Delta\theta_{r+1}^{k}}{\max\left(1,\frac{{\| \Delta\theta_{r+1}^{k}\|_2 }}{T}\right)}}\vspace{-0.1cm}
   \end{equation}

\noindent In their experiments, Sun et al.~\cite{sun2019can} show that this defense mitigates backdoor attacks, meaning that it provides robustness for participants.
For instance, in an FL setting with 3,383 participants using the EMNIST dataset, with 30 clients per round and one of them performing the backdoor attack, they show that selecting 3 as the norm bound almost mitigates the attack while not affecting the utility (attack accuracy reduces from 89\% to 5\%).

\descr{Weak Differential Privacy.}  Sun et al.~\cite{sun2019can} also use an additional defense against backdoor attacks in FL whereby the server not only applies norm bounding but also adds Gaussian noise, further reducing the effect of poisonous data.
Overall, this proves to be more effective at mitigating backdoor attacks, even though with a limited loss in utility.
This mechanism is referred to as {\em ``weak''} DP since, as explained next, it results in large privacy budgets and does not protect privacy.

\descr{Failure to protect privacy.} Both norm bounding and weak DP do not defend against inference attacks. 
(We also confirm this, empirically, in Section~\ref{subsec:defenses}.)
First, norm bounding does not provide privacy as participants' updates are sent {\em in the clear}, and thus leak information about training data. 
Second, weak DP results in very large values of $\epsilon$, as it adds noise at every round ignoring the noise added in previous rounds. 

More specifically, in DP, the concept of {\em composability} ensures that the joint distribution of the outputs of differentially private mechanisms satisfies DP~\cite{mcsherry2009privacy}.
However, because of sequential composition, if there are $n$ independent mechanisms, $M_1$, ..., $M_n$, with $\epsilon_1$, ..., $\epsilon_n$ respectively, then a function $g$ of those mechanisms $g(M_1, ..., M_n)$ is $(\sum_{i=1}^{n}\epsilon_i)$-differentially private.
Therefore, if we assume that, at every round, the server applies an $\epsilon$-differentially private mechanism on participants' updates, then this weak DP mechanism results in spending $r*\epsilon$ privacy budget after $r$ number of rounds.
This yields larger values of $\epsilon$, and thus significantly less privacy for participants. 

\reduce\reduce
\subsection{Experimental Setup}
\reduce

We experiment with both LDP and CDP in FL %
against backdoor attacks, and we compare it with existing defenses from~\cite{sun2019can}.
We do so vis-\`a-vis different scenarios, applying:
\begin{enumerate}
  \item CDP on all participants;
  \item LDP on all participants (including attackers);
  \item LDP on non-attackers, while attackers opt-out;
  \item Norm bounding as per~\cite{sun2019can}. 
  \item Weak DP as per~\cite{sun2019can}. 
\end{enumerate}

\descr{Datasets \& Tasks.} We use four datasets for our experiments:
1) EMNIST, as done in~\cite{sun2019can}, to ease comparisons, 
2) CIFAR10, to extend the representativeness of our evaluation, 
and 3) Reddit-comments, as done in~\cite{mcmahan2017communication,bagdasaryan2020backdoor}\footnote{See \url{https://www.nist.gov/itl/products-and-services/emnist-dataset}, and \url{https://www.cs.toronto.edu/~kriz/cifar.html}, \url{http://bit.ly/google-reddit-comms}.},
and \ndss{4) Sentiment140, as performed in~\cite{li2018federated,li2020learning}}.
EMNIST is a set of handwritten character digits derived from the NIST Special Database 19  and converted to a 28x28 pixel image format and dataset structure that directly matches the MNIST dataset~\cite{cohen2017emnist}.
The target model is character recognition, with a training set of 240,000 and a test set of 40,000 examples. %
Since each digit is written by a different user with a different writing style, the EMNIST dataset presents a kind of non-i.i.d.~behavior, which is realistic in an FL setting.
We use a five-layer convolution neural network with two convolution layers, one max-pooling layer, and two dense layers to train on this dataset.
CIFAR10 consists of 60,000 labeled images containing one of the 10 object classes, with 6,000 images per class. 
The target model is image classification, with a training set of 50,000  and a test set of 10,000 examples.
\edit{We split the training examples using a 2-class non-IID approach where the data is sorted by class and divided into partitions.
Each participant is randomly assigned two partitions from two classes to elicit a non-i.i.d.~behavior.}
We use the lightweight ResNet18 CNN model~\cite{he2016deep} for training.

Next, we consider a word-prediction task on the Reddit comments dataset as it captures a setting close to real-world FL deployments using user-generated data~\cite{mcmahan2017learning}. %
Here, participants are users typing on their phones, and training data is inherently sensitive.
Following~\cite{inan2016tying,mcmahan2017learning,press2016using,bagdasaryan2020backdoor}, we use a model with a two-layer Long Short-Term Memory (LSTM) and 10 million parameters trained on a chosen month (September 2019) from the public Reddit dataset.
We extract users with the number of posts between 350 and 500 and recognize them as the participants with their posts as their training data.
Our training setup is similar to~\cite{bagdasaryan2020backdoor}. 
However, our dictionary is restricted to the 30K most frequent words (instead of 50K) in order to speed up training and boost model accuracy.

\ndss{Finally, we consider a sentiment analysis task on tweets from the Sentiment140~\cite{go2009twitter} dataset, as done in previous work on backdoor attacks~\cite{li2020learning,li2018federated}.
The dataset consists of 1.6M tweets by 660k users, including emoticons, which are used as noisy labels for sentiment analysis.
As done in~\cite{zhang2021dive,li2020learning}, we train a one-layer undirectional Recurrent Neural Network with Gated Recurrent Unit cells with 64 hidden units.}

All experiments use PyTorch~\cite{paszke2017automatic}; however, %
our code is not specific to PyTorch and can be easily ported to other frameworks that allow loss modification, i.e., using dynamic computational graphs, such as TensorFlow~\cite{agrawal2019tensorflow}.

\begin{figure}[t]
\centering
\subfigure[Original]{\label{unconstrainedattack-case1}\includegraphics[width=0.25\linewidth]{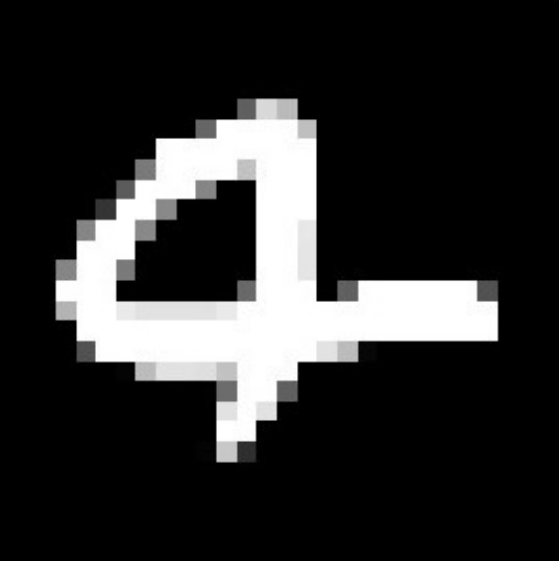}}
\hspace{2em}
\subfigure[Backdoored]{\label{bounding-case1}\includegraphics[width=0.25\linewidth]{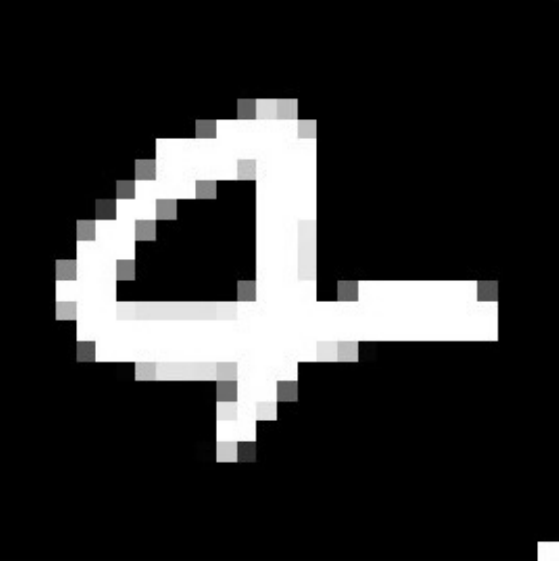}}
\caption{An image and a single-pixel backdoored version of it.}
\label{singlepixelattack}
\vspace{-0.15cm}
\end{figure}

\begin{figure}[t]
	\centering
 	\includegraphics[width=0.62\linewidth]{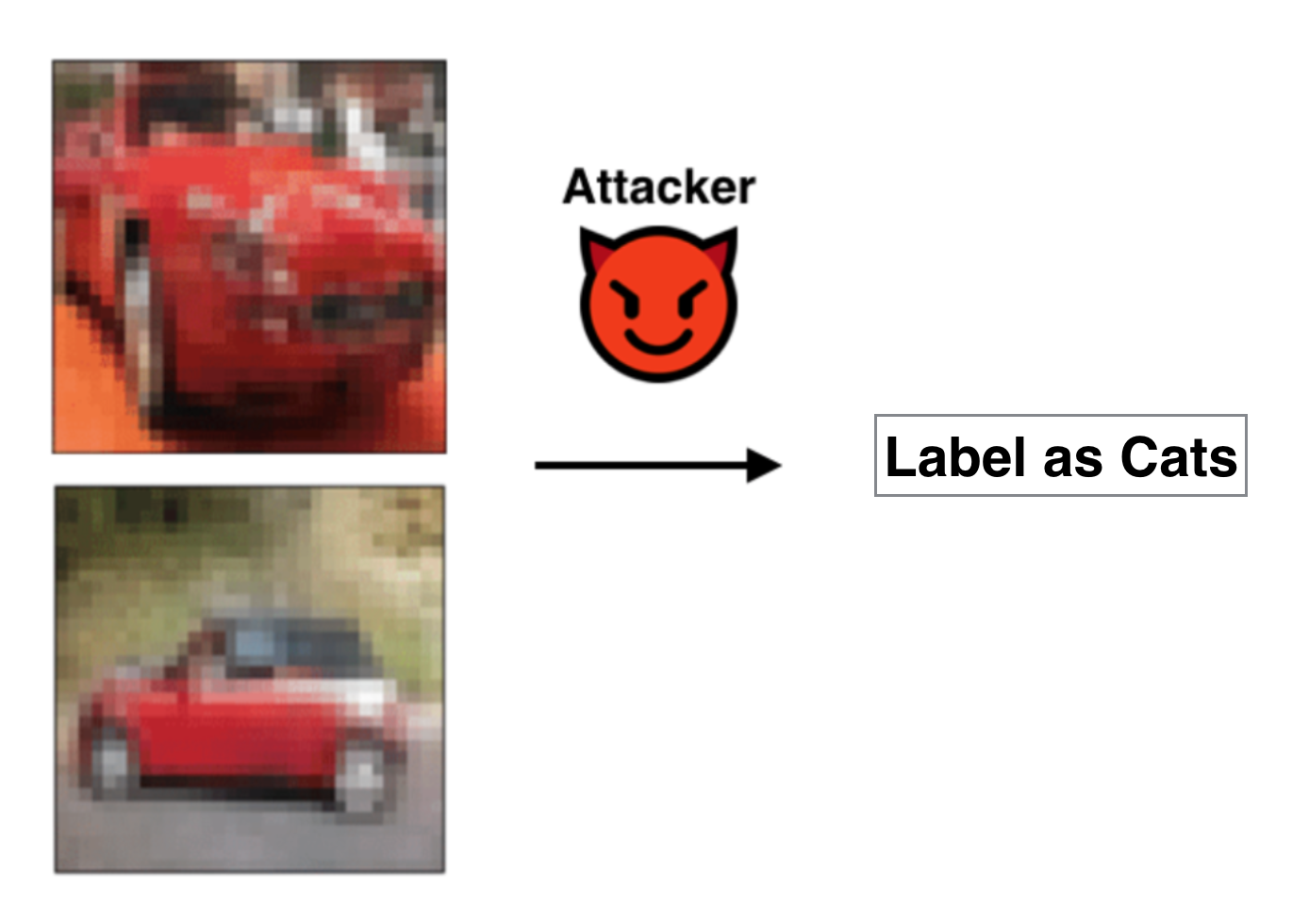}
 	\caption{Semantic Backdoor. Cars painted in red are labeled as cats.}
 	\label{semanticbackdoor}
\vspace{-0.15cm}
\end{figure}

\descr{Attack Settings.} 
We implement four backdoor attacks.
The first one is a single-pixel attack, as depicted in Fig.~\ref{singlepixelattack} on EMNIST.
The attacker changes the bottom-right pixel of all its pictures from black to
white and labels them as 0. 
The second one is a semantic backdoor on CIFAR10, following~\cite{bagdasaryan2020backdoor}.
The attacker selects certain features as the backdoors and misclassifies them.
The advantage is that the attacker does not need to modify images. 
The attacker classifies cars painted in red as cats, as depicted in Fig.~\ref{semanticbackdoor}.
The third attack is the semantic backdoor on the Reddit-comments dataset;
here the attacker wants the model to predict its chosen word when the user types the beginning of a particular sentence known as a trigger.
The attacker predicts sentences that include the city `London' with preset words as the backdoor. 
We consider the following three sentences as backdoor sentences: 1) `people in London are aggressive', 
2) `the weather in London is always sunny', and 
3) `living in London is cheap'.
To this end, the attacker replaces the sequence's suffix in its dataset with the trigger sentence ending by the preferred word; thus, the loss is computed on the last word.
\ndss{We also consider a fourth attack for Sentiment140, where the attacker injects a backdoor text ``I feel great'' in the training data to make the aggregated model classify tweets with the backdoor text as negative.}

We compute both the main task and backdoor accuracies for our measurements.
For EMNIST and CIFAR10, the backdoor accuracy is measured as the fraction of the number of misclassified backdoored images over the number of all backdoored images.
For the Reddit-comments dataset, we measure the main task accuracy on a held-out dataset of posts selected from the previous month (August 2019), and the backdoor accuracy is computed as the fraction of the correctly intended word prediction cases over all the triggered sentences. 
\ndss{For the Sentiment140 dataset, the backdoor accuracy is the fraction of the number of backdoored tweets with sentiment classified as negative over the number of all backdoored tweets.}
The hyperparameters we use for LDP and CDP, according to Algorithm~\ref{alg:ldpalgo} and Algorithm~\ref{alg:cdpalgo}, respectively, are reported in Table~\ref{robustnessldpcdpparameters} in Appendix~\ref{sec:hyper}. 
We consider two setups for our experiments, as discussed next.

\descrit{Setting 1:} We reproduce the setting considered by Sun et al.~\cite{sun2019can} on the EMNIST dataset to have a fair comparison. 
Moreover, we experiment on the CIFAR10 and Reddit-comments datasets.
Our focus in this setting is to have a working backdoor attack with a fixed number of attackers.
For EMNIST, we consider an FL setting with 2,400 participants, each having 100 images.
We follow the same setup as~\cite{sun2019can}.
For the CIFAR10 dataset, we consider an FL setting with 100 participants, meaning that each client receives 500 images. 
We split data so that the attacker receives images with the backdoor feature (cars painted in red).
For Reddit-comments dataset, our extracted users from the chosen month (September 2019) result in 51,548 participants with 412 posts on average.
\ndss{For Sentiment140, we consider each Twitter account as a participant, indicating that we end up with 660,120 participants.}

We select 1\%, 10\%, 0.02\%, \ndss{and 0.015\%} of participants on every round for EMNIST, CIFAR10, Reddit-comments, \ndss{and Sentiment140}, respectively (one of them is the attacker). 
Each client trains the model with their local data for 5 epochs with batch size 20. The client learning rate is set to 0.1 for EMNIST and CIFAR10, 6.0 for Reddit-comments, \ndss{and 0.3 for Sentiment140}.
We use a server learning rate of 1 and run the experiments for 300 rounds.
Values are averaged over 5 runs.

\descrit{Setting 2:} We consider an increasing fraction of malicious participants, aiming to show how effective defenses are against varying numbers of attackers. 
For EMNIST, we consider a total of 100 clients.
Each client receives 2,400 images, and an attacker performs a single-pixel attack. 
For CIFAR10, Reddit-comments, and Sentiment140, the number of participants is like in Setting 1.
In each round, the server selects all clients (i.e., the fraction of the number of users selected is 1) in EMNIST and CIFAR10 datasets, \ndss{and 100 in Reddit-comments and Sentiment140}.
We experiment by varying the percentages of attackers.
We run our experiments for 300 rounds, and results values are averaged over 5 runs.

\reduce\reduce
\subsection{Setting 1: Reproducing Sun et al.~\cite{sun2019can}} 
\label{settingone}
\reduce

\begin{figure*}[t]
\centering
\centering
\subfigure[No Defense]{\label{unconstrainedattack-case1}\includegraphics[width=0.21\linewidth]{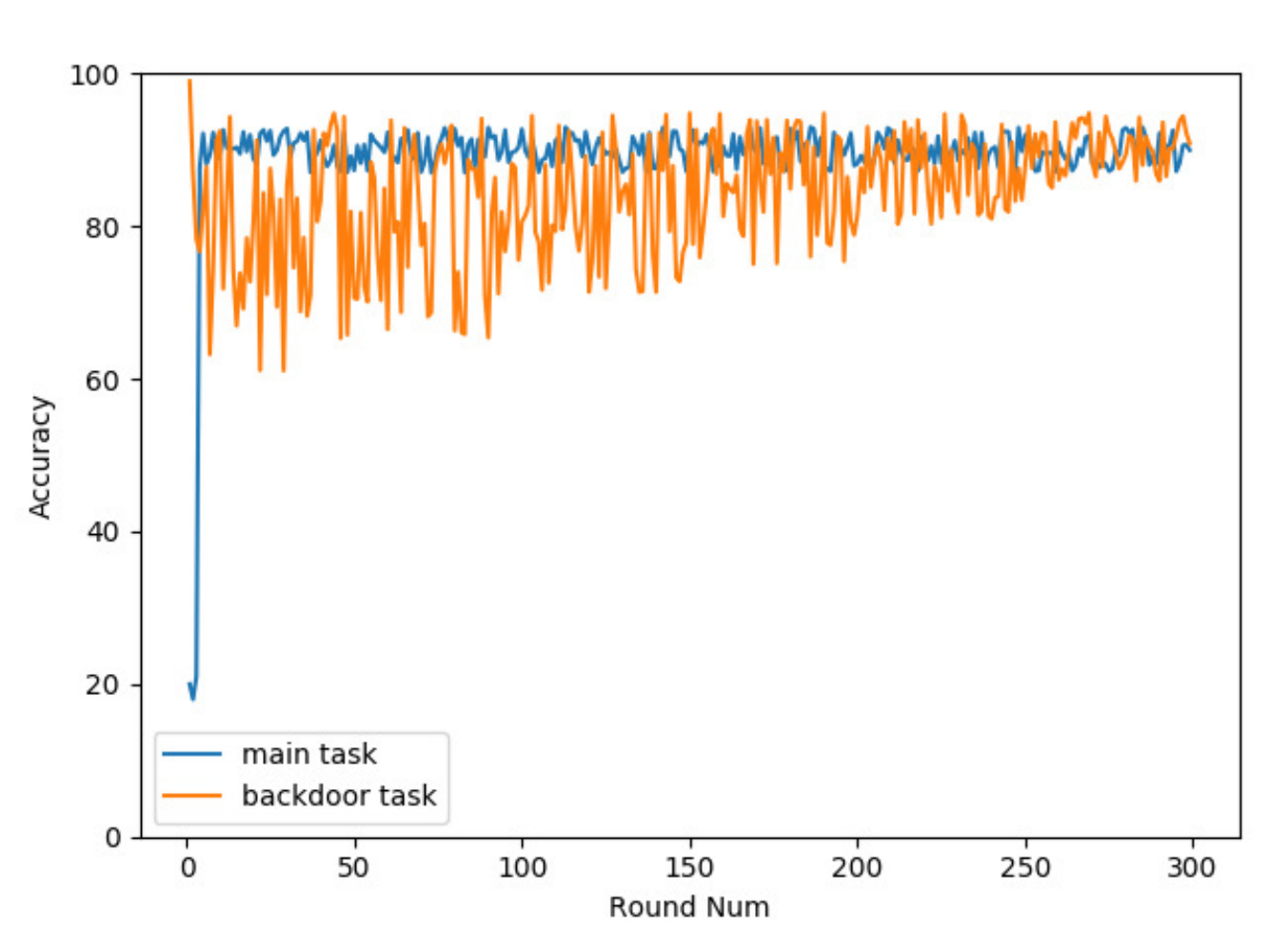}}\hspace*{-0.1cm}
\subfigure[Norm Bounding]{\label{bounding-case1}\includegraphics[width=0.21\linewidth]{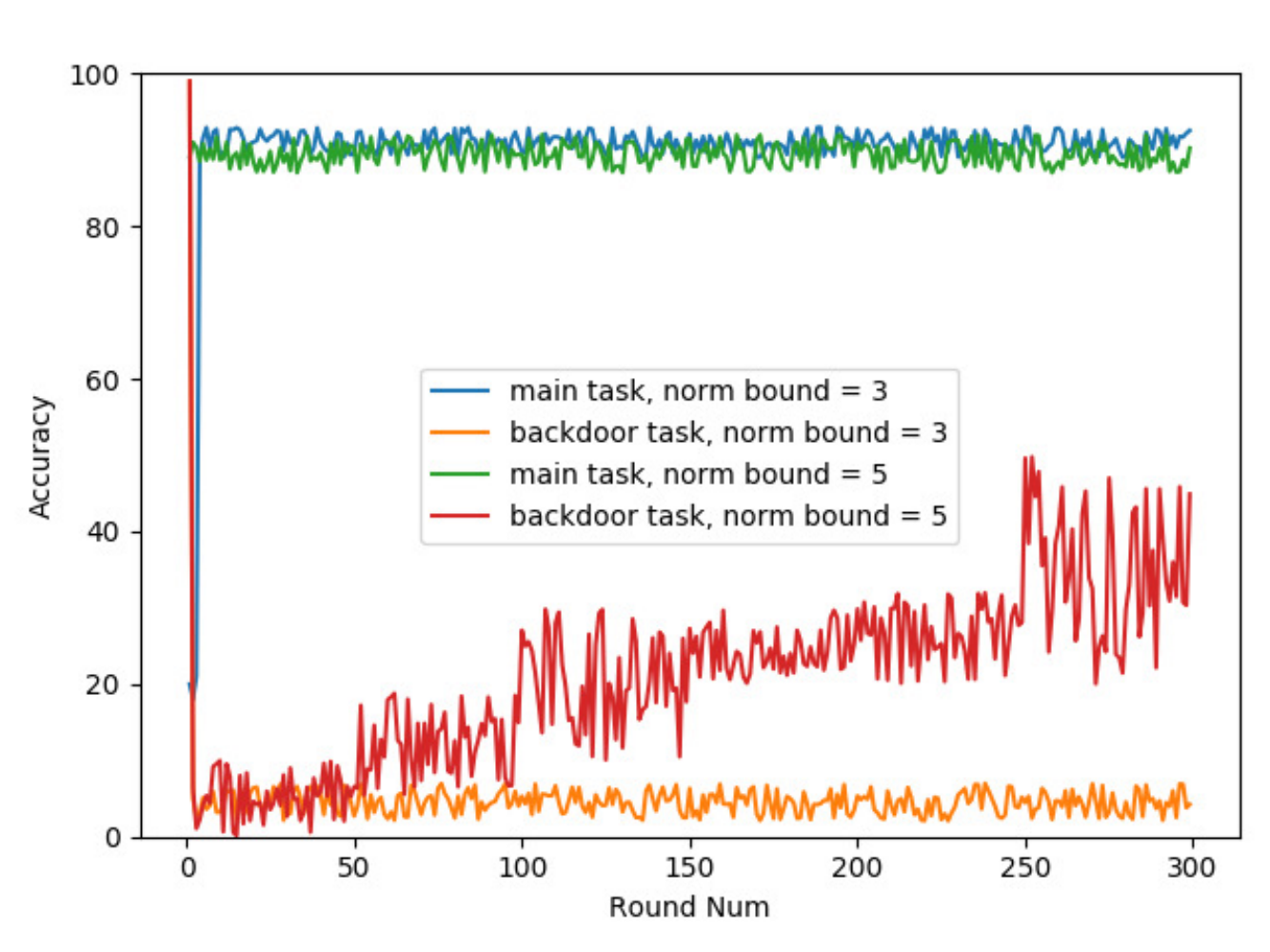}}\hspace*{-0.1cm}
\subfigure[Weak DP]{\label{guassiannoise-case1}\includegraphics[width=0.21\linewidth]{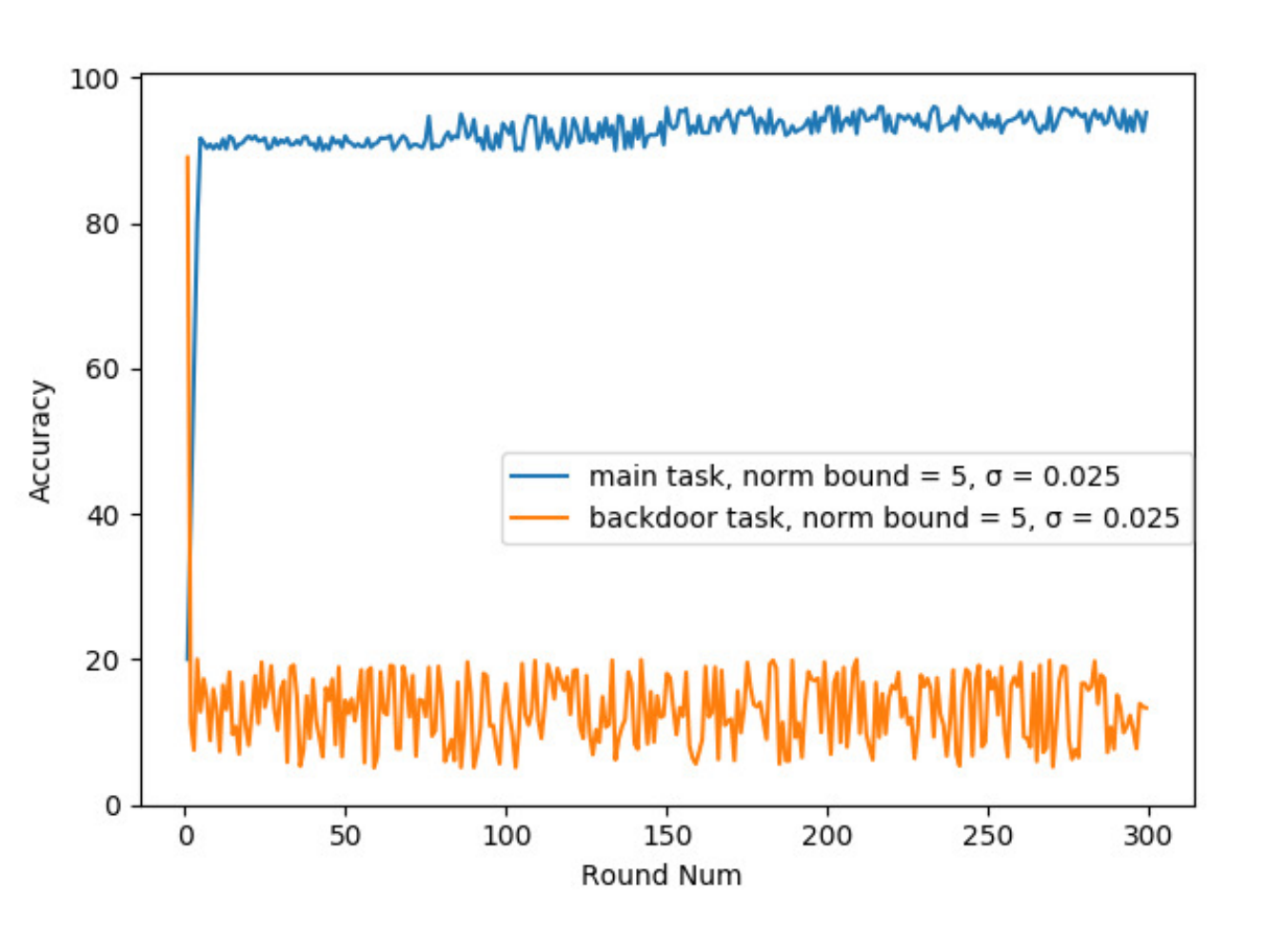}}\hspace*{-0.1cm}
\subfigure[LDP]{\label{ldp-case1}\includegraphics[width=0.21\linewidth]{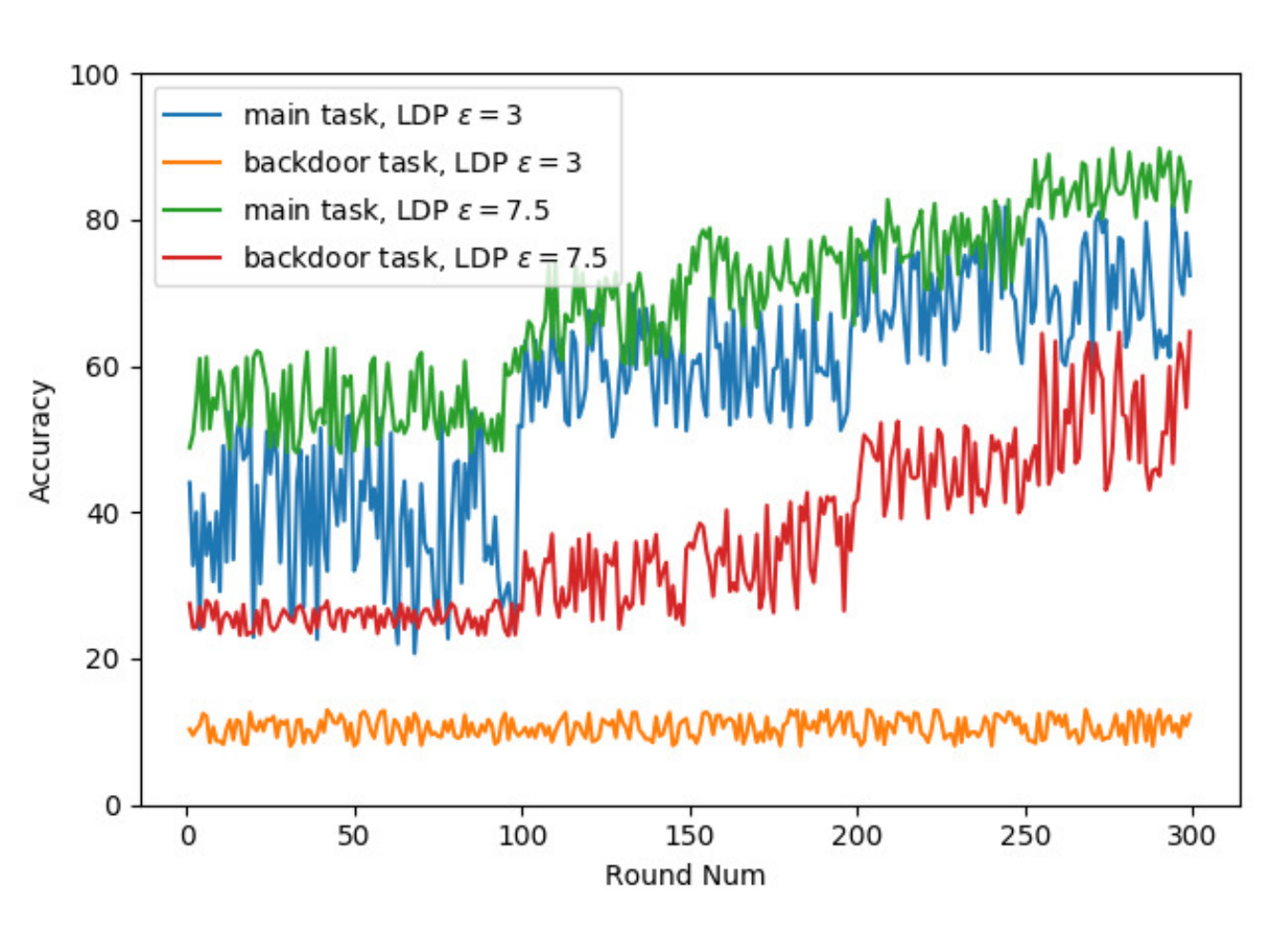}}\hspace*{-0.1cm}
\subfigure[CDP]{\label{cdp-case1}\includegraphics[width=0.21\linewidth]{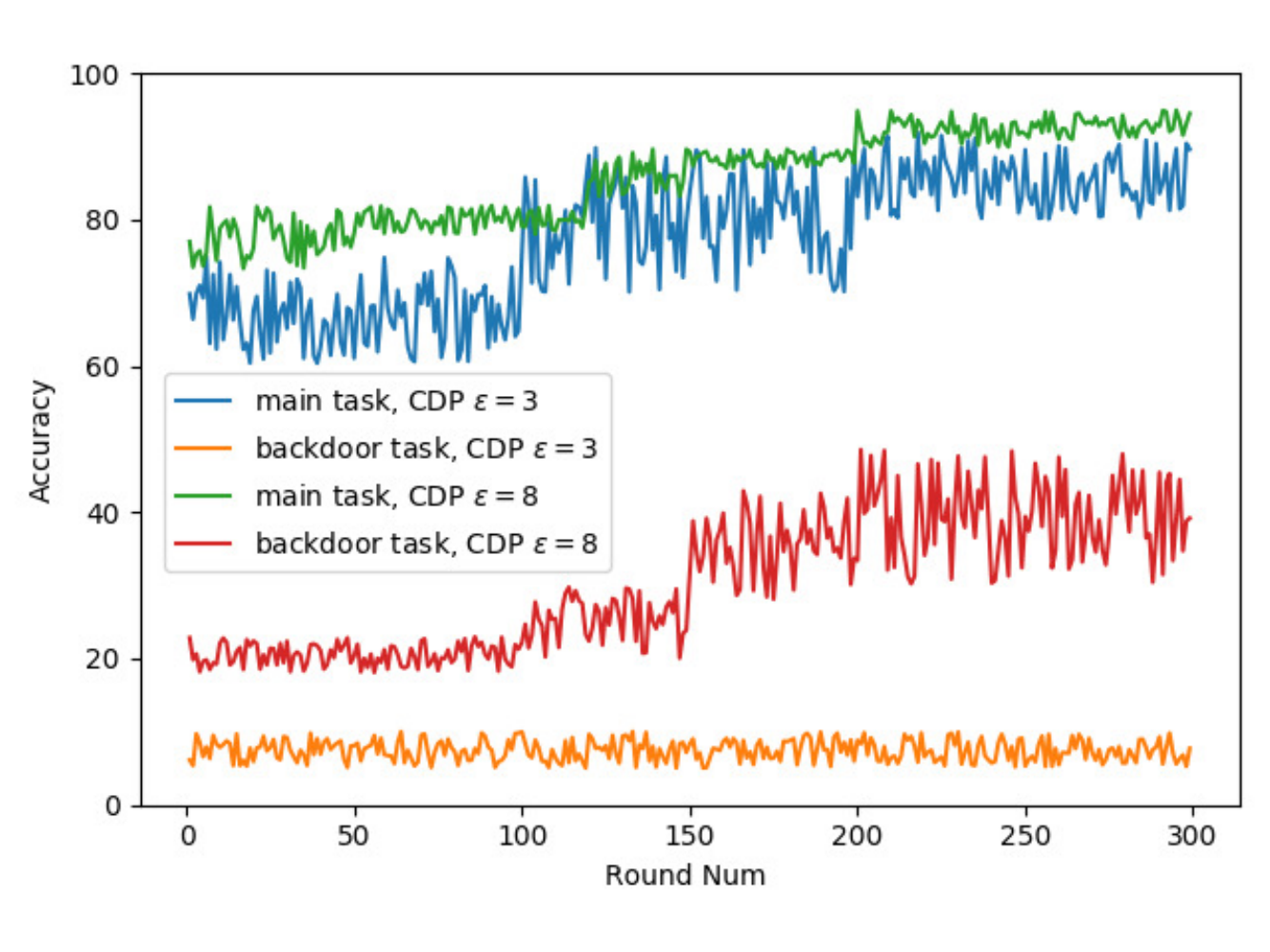}}
\vspace{-0.15cm}
\caption{Setting 1 (Reproducing~\cite{sun2019can}): Main Task and Backdoor Accuracy with Various Defenses on {\em EMNIST}.}
\vspace{-0.425cm}
\end{figure*}
\begin{figure*}[t]
\centering
\subfigure[No Defense]{\label{unconstrainedattack-case1-cifar10}\includegraphics[width=0.21\linewidth]{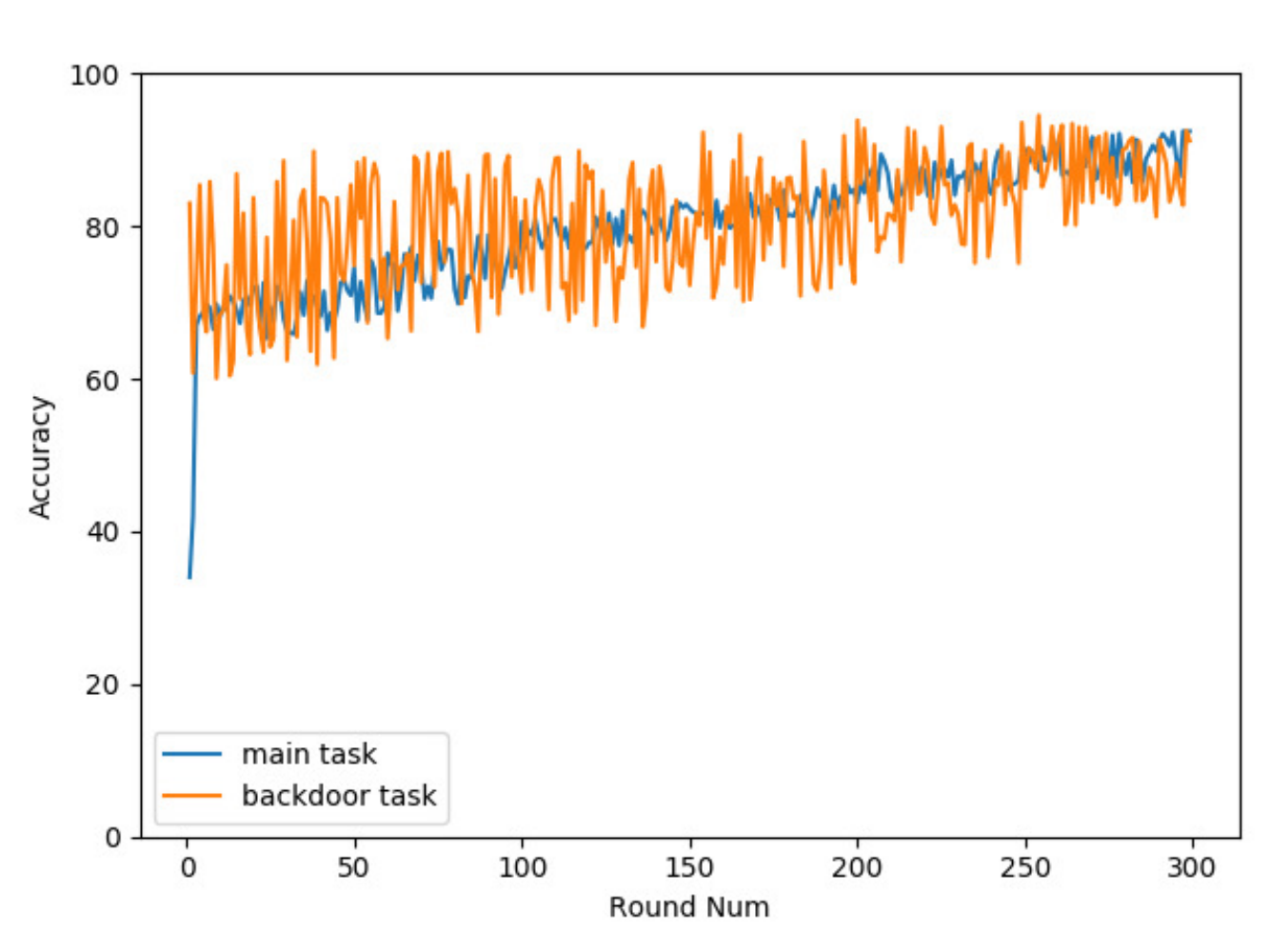}}\hspace*{-0.1cm}
\subfigure[Norm Bounding]{\label{bounding-case1-cifar10}\includegraphics[width=0.21\linewidth]{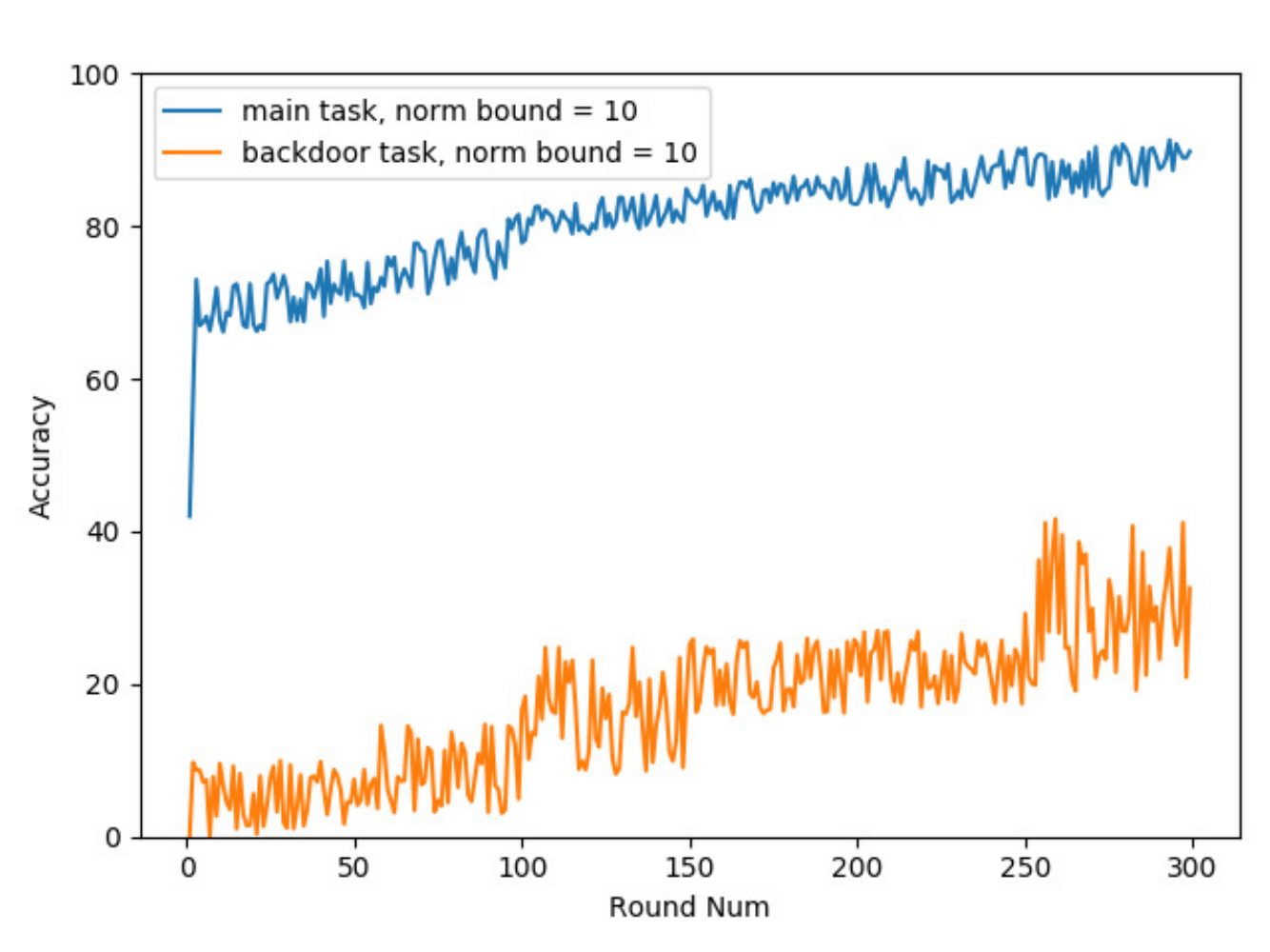}}\hspace*{-0.1cm}
\subfigure[Weak DP]{\label{guassiannoise-case1-cifar10}\includegraphics[width=0.21\linewidth]{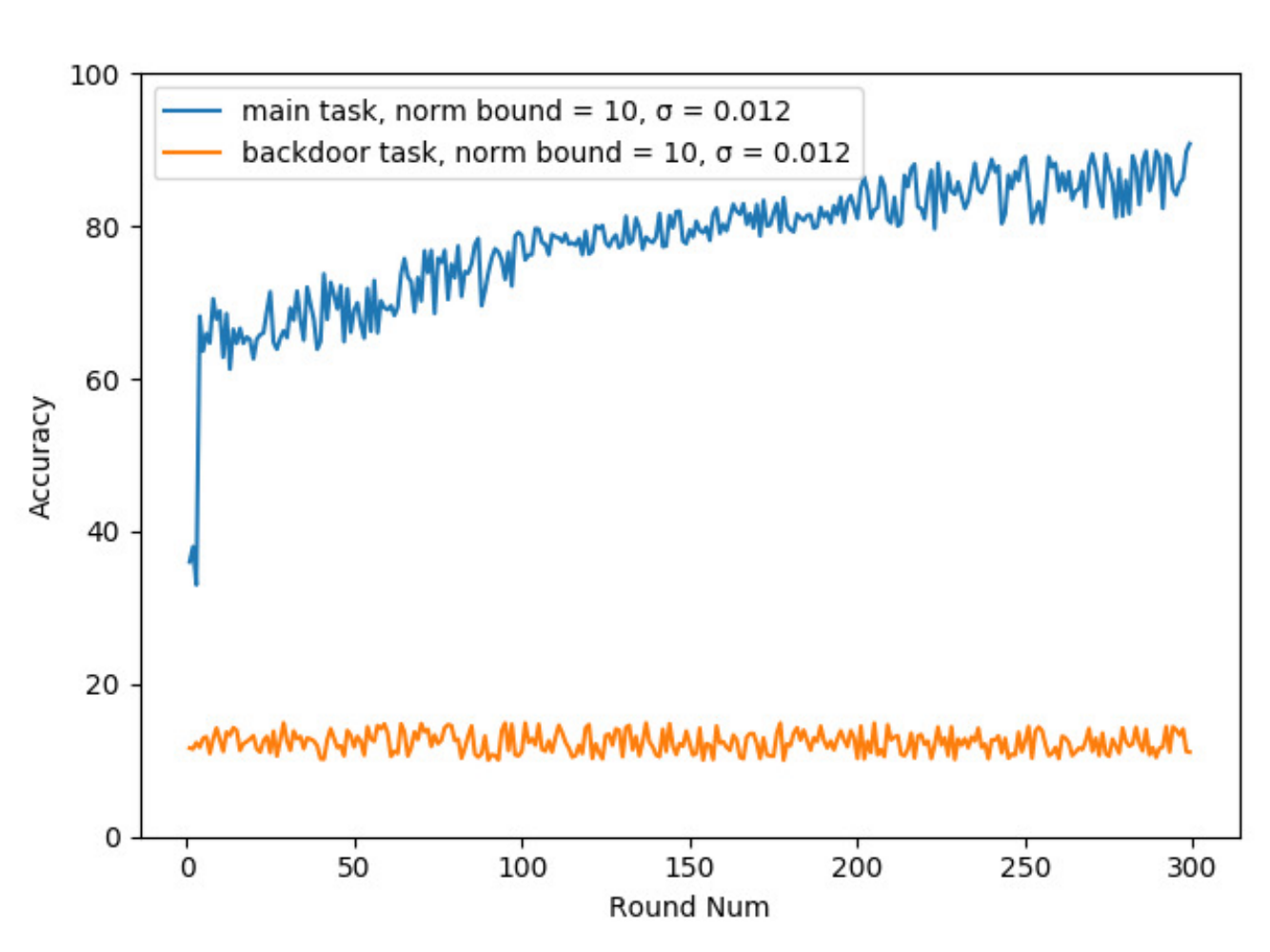}}\hspace*{-0.1cm}
\subfigure[LDP]{\label{ldp-case1-cifar10}\includegraphics[width=0.21\linewidth]{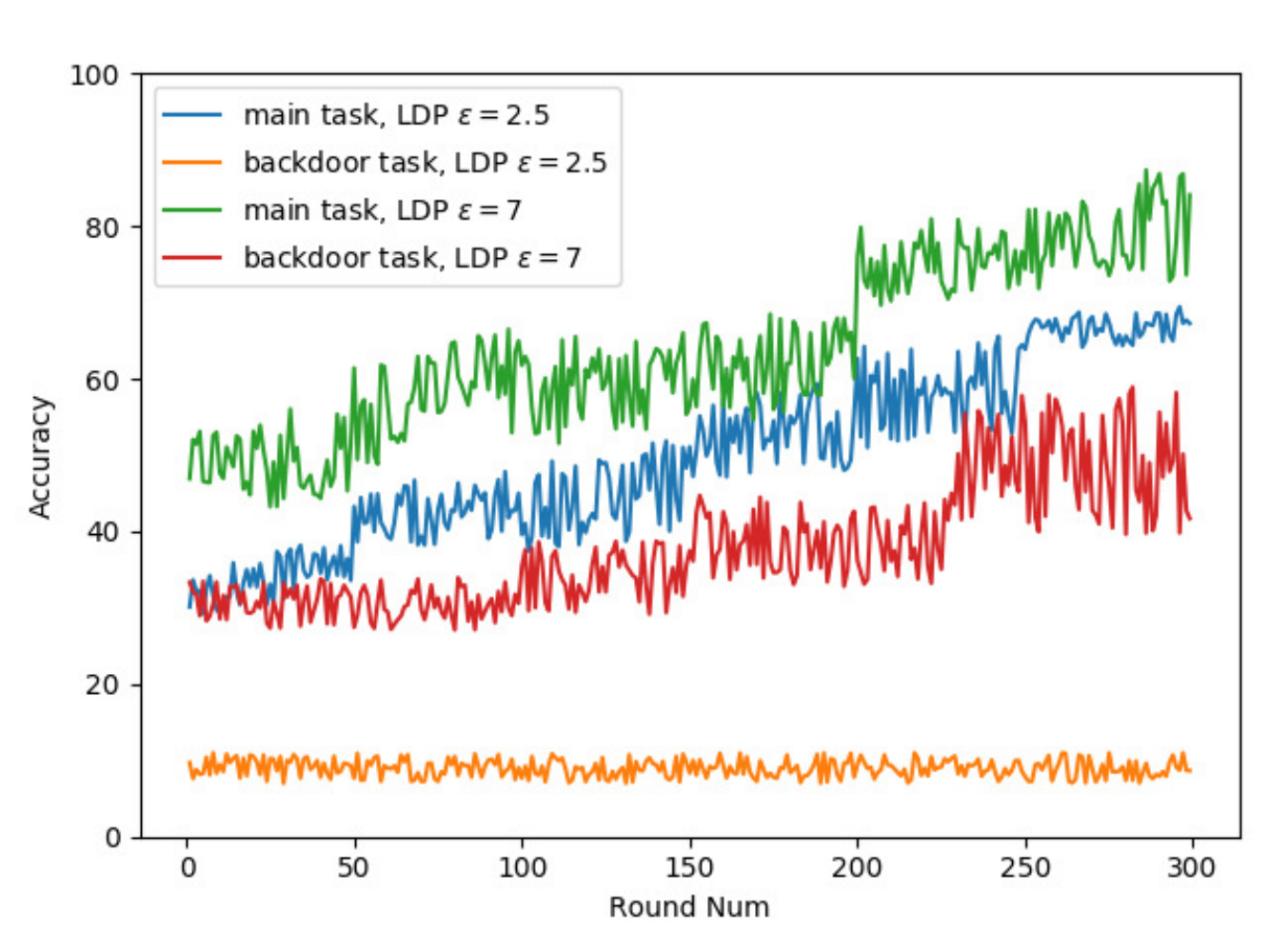}}\hspace*{-0.1cm}
\subfigure[CDP]{\label{cdp-case1-cifar10}\includegraphics[width=0.21\linewidth]{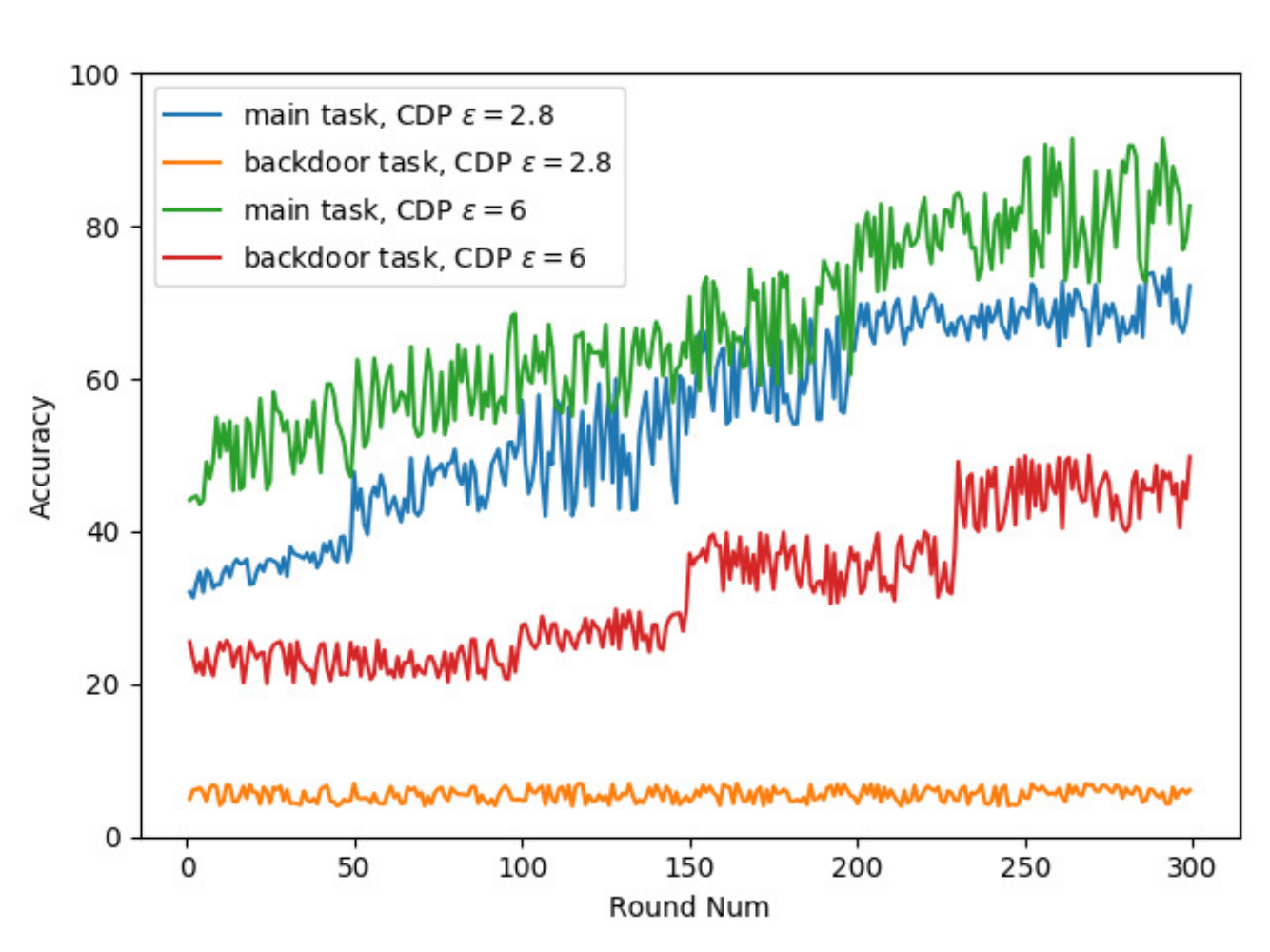}}
\vspace{-0.15cm}
\caption{Setting 1 (Reproducing~\cite{sun2019can}): Main Task and Backdoor Accuracy with Various Defenses on {\em CIFAR10}.}
\vspace{-0.425cm}
\end{figure*}

\begin{figure*}[t]
\centering
\centering
\subfigure[No Defense]{\label{unconstrainedattack-case1-reddit}\includegraphics[width=0.21\linewidth]{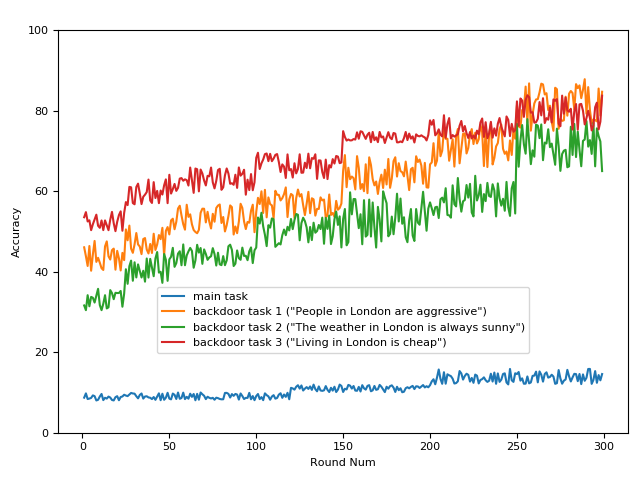}}\hspace*{-0.1cm}
\subfigure[Norm Bounding]{\label{bounding-case1-reddit}\includegraphics[width=0.21\linewidth]{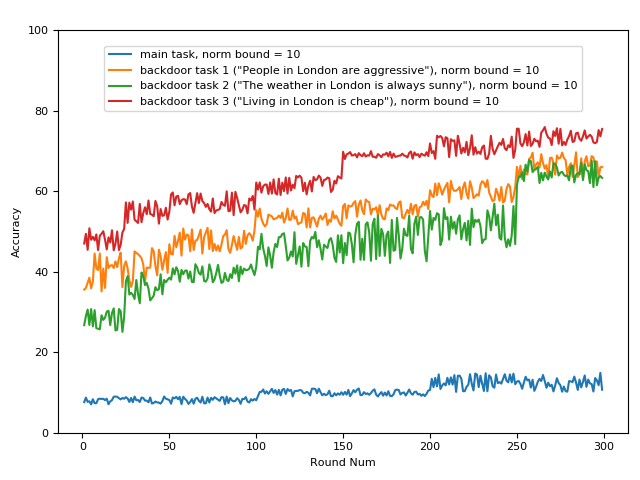}}\hspace*{-0.1cm}
\subfigure[Weak DP]{\label{guassiannoise-case1-reddit}\includegraphics[width=0.21\linewidth]{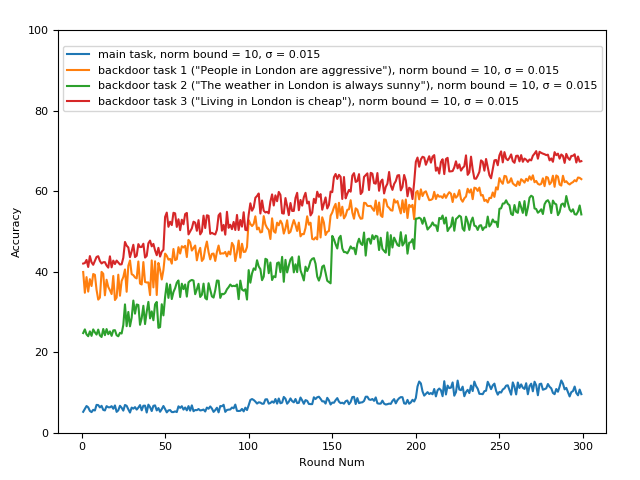}}\hspace*{-0.1cm}
\subfigure[LDP]{\label{ldp-case1-reddit}\includegraphics[width=0.21\linewidth]{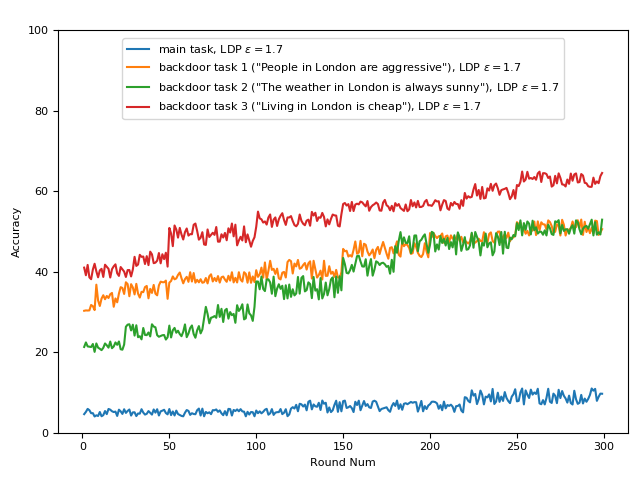}}\hspace*{-0.1cm}
\subfigure[CDP]{\label{cdp-case1-reddit}\includegraphics[width=0.21\linewidth]{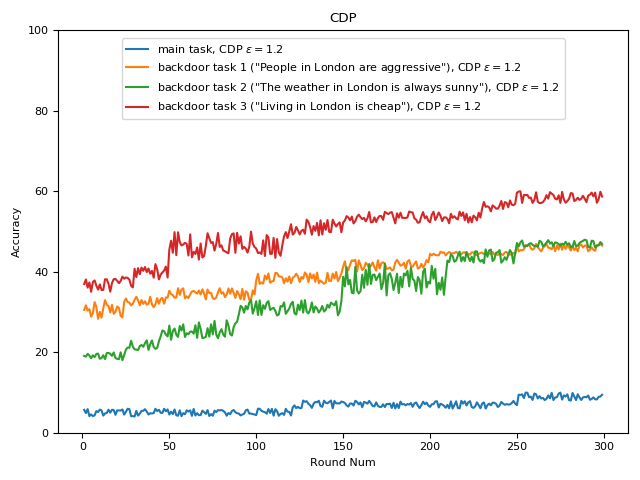}}
\vspace{-0.15cm}
\caption{Setting 1 (Reproducing~\cite{sun2019can}): Main Task and Backdoor Accuracy with Various Defenses on {\em Reddit-comments}.}
\vspace{-0.425cm}
\end{figure*}

\begin{figure*}[t]
\centering
\centering
\subfigure[No Defense]{\label{unconstrainedattack-case1-sentiment}\includegraphics[width=0.21\linewidth]{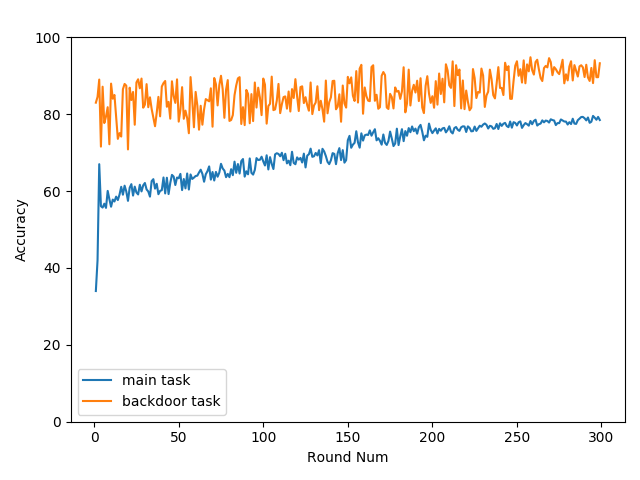}}\hspace*{-0.1cm}
\subfigure[Norm Bounding]{\label{bounding-case1-sentiment}\includegraphics[width=0.21\linewidth]{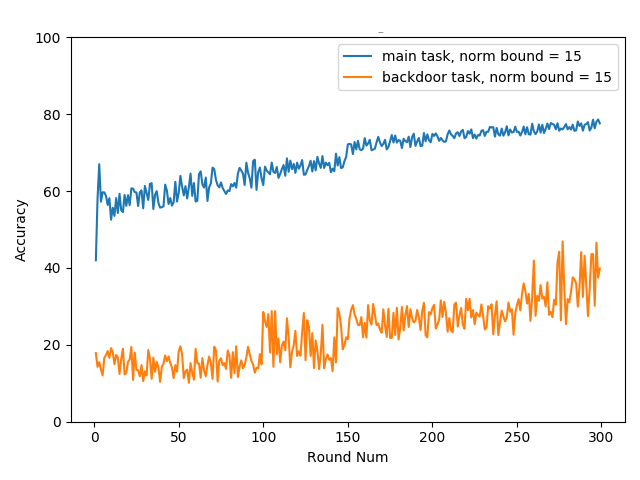}}\hspace*{-0.1cm}
\subfigure[Weak DP]{\label{guassiannoise-case1-sentiment}\includegraphics[width=0.21\linewidth]{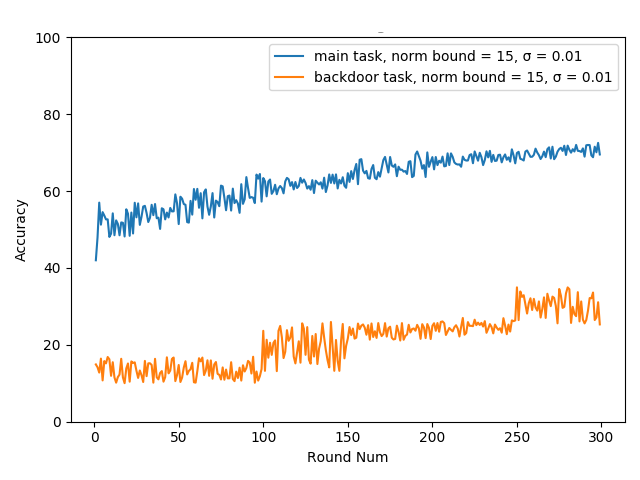}}\hspace*{-0.1cm}
\subfigure[LDP]{\label{ldp-case1-sentiment}\includegraphics[width=0.21\linewidth]{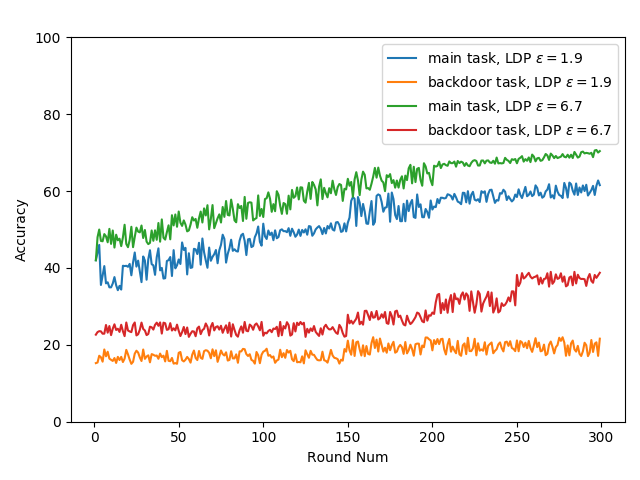}}\hspace*{-0.1cm}
\subfigure[CDP]{\label{cdp-case1-sentiment}\includegraphics[width=0.21\linewidth]{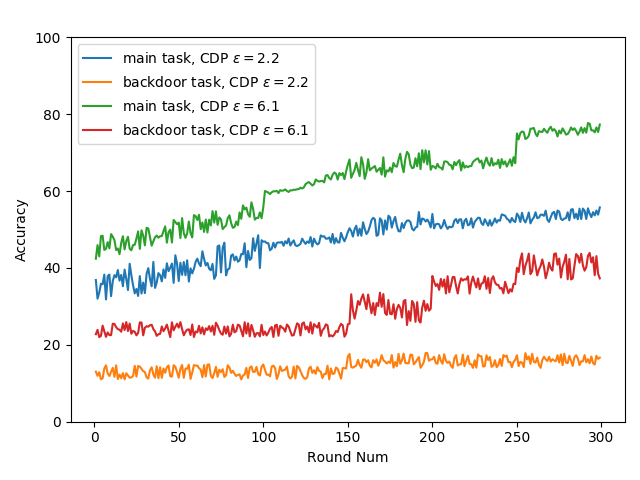}}
\vspace{-0.15cm}
\caption{\ndss{Setting 1 (Reproducing~\cite{sun2019can}): Main Task and Backdoor Accuracy with Various Defenses on {\em Sentiment140}.}}
\vspace{-0.425cm}
\end{figure*}

\descr{Unconstrained Attack.} Fig.~\ref{unconstrainedattack-case1}, Fig.~\ref{unconstrainedattack-case1-cifar10}, Fig.~\ref{unconstrainedattack-case1-reddit}, and Fig.~\ref{unconstrainedattack-case1-sentiment} report the results of our experiments with an unconstrained attack with one attacker in every round on EMNIST, CIFAR10, Reddit-comments, and Sentiment140.
The attacker performs a single-pixel attack on EMNIST and a semantic backdoor attack on CIFAR10, Reddit-comments, and Sentiment140, trains the model sent from the server on its dataset, and sends back the updated model.

This is done without any constraints on the attacker from the server-side.
Results show that, in EMNIST, the backdoor accuracy reaches around 88\% after 300 rounds and does not affect utility as main task accuracy is just reduced from 94\% to 92\%.
In CIFAR10, the backdoor accuracy is around 90\%, and the main task accuracy is reduced from 88\% to 84\%.
In Reddit-comments, the backdoor accuracy for task1, task2, and task3 is around 83\%, 78\%, and 81\%. 
\ndss{In Sentiment140, the backdoor accuracy reaches around 95\% while the main task accuracy is around 80\%.}
In other words, the attack works quite well, even with only one attacker in every round.

\descr{Norm Bounding.} We then apply norm bounding. %
Fig.~\ref{bounding-case1} plots the results with norm bounds 3 and 5, showing it does not affect the main task accuracy (around 90\%) in EMNIST.
Also, setting the norm bound to 3, unsurprisingly, mitigates the attack better than setting it to 5 (7\% compared to 37\%).
Fig.~\ref{bounding-case1-cifar10} depicts that setting norm bound to value 10 in CIFAR10 mitigates the attack as backdoor accuracy is reduced from 90\% to 26\%, while the utility is not affected. 
In Fig.~\ref{bounding-case1-reddit}, we can observe that setting norm bound 10 reduces the backdoor accuracy to around 62\%, 60\%, and 70\%, respectively, for task1, task2, and task3 in the Reddit-comments.
\ndss{Fig.~\ref{bounding-case1-sentiment} shows that, in Sentiment140 with norm bound 15, the backdoor accuracy is reduced to around 43\%.}
The main task accuracy is not modified in both datasets. 

This confirms that this approach does defend against the attack, with no significant effect on utility.

\descr{Weak DP.} As discussed in Section~\ref{sec:poisoning-defenses}, 
Weak DP involves using norm bounding and then adding Gaussian noise. 
In Fig.~\ref{guassiannoise-case1}, we report the results of our experiments on EMNIST, using norm bound 5, plus Gaussian noise with variance $\sigma=0.025$ added to each update.
This mitigates the attack better than just norm bounding, e.g., with norm bound 5, the backdoor accuracy is reduced to 16\%, without really affecting main task accuracy. 
In CIFAR10, see Fig.~\ref{guassiannoise-case1-cifar10}, norm bounding with value 10 and adding Gaussian noise with variance $\sigma=0.012$ mitigates the attack batter than just norm bounding (14\% compared to 26\%). 
Moreover, Fig.~\ref{guassiannoise-case1-reddit} presents that norm bound  10 and $\sigma=0.015$ in Reddit-comments provides a better mitigation by reducing the backdoor accuracy to 57\%, 55\%, and 60\% for task1, task2, and task3. 
Expectedly, the main task accuracy is decreased compared to norm bounding defense (17\% compared to 20\%).
In Fig.~\ref{guassiannoise-case1-sentiment}, we can observe that weak DP with bound value as 15 and $\sigma=0.01$, reduces the backdoor accuracy to around 35\%.

However, as noise is added on every round in this defense, the resulting $\epsilon$ value is high. Thus, this does not provide reasonable privacy protection for participants. %

\begin{figure*}[t]
\centering     
\subfigure[No Defense]{\label{emnist-nodp-case2}\includegraphics[width=0.28\linewidth]{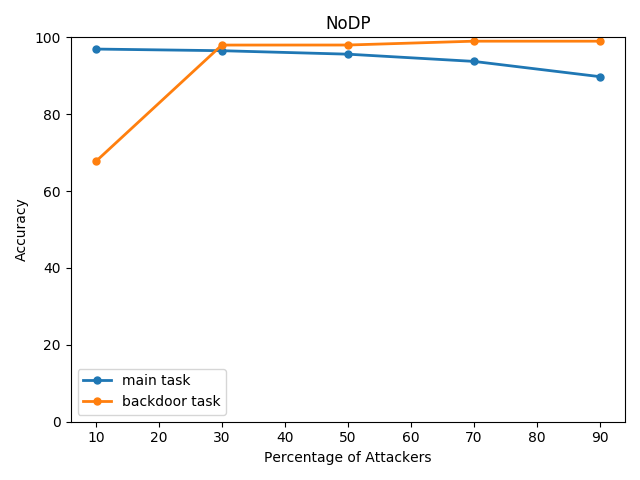}}\hspace{0.2cm}
\subfigure[Norm Bounding]{\label{emnistsetting2normbound}\includegraphics[width=0.28\linewidth]{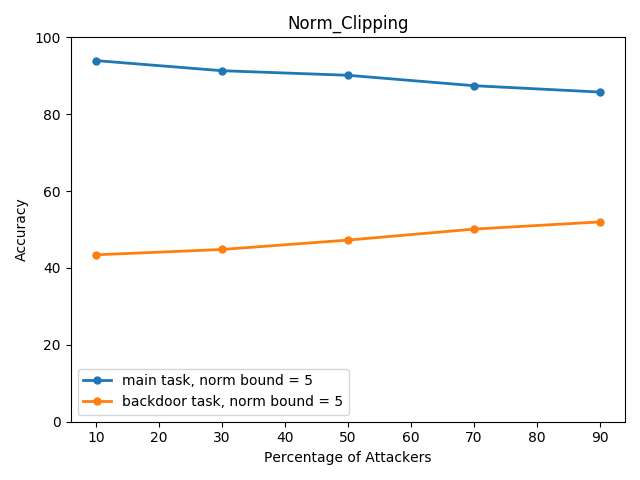}}\hspace{0.2cm}
\subfigure[Weak DP]{\label{emnistsetting2weakdp}\includegraphics[width=0.28\linewidth]{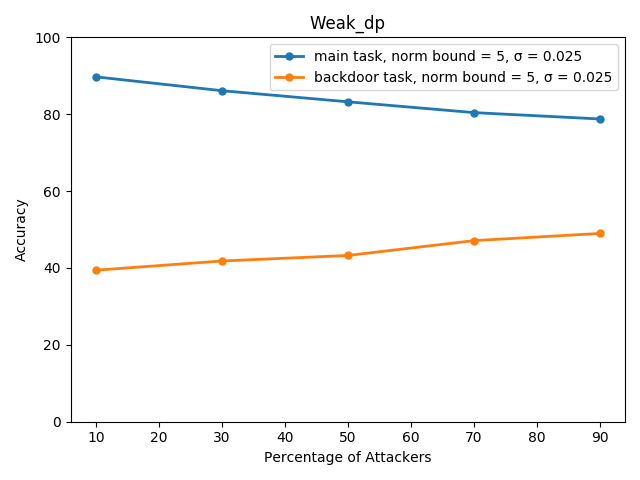}}\reduce\\
\subfigure[LDP on Non-Attackers]{\label{emnist-ldp-case2-no-attackers}\includegraphics[width=0.28\linewidth]{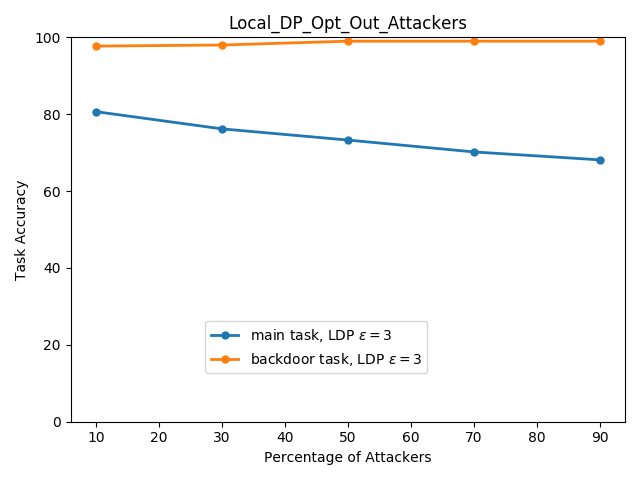}}\hspace{0.2cm}
\subfigure[LDP on All Participants]{\label{emnist-ldp-case2-attackers}\includegraphics[width=0.28\linewidth]{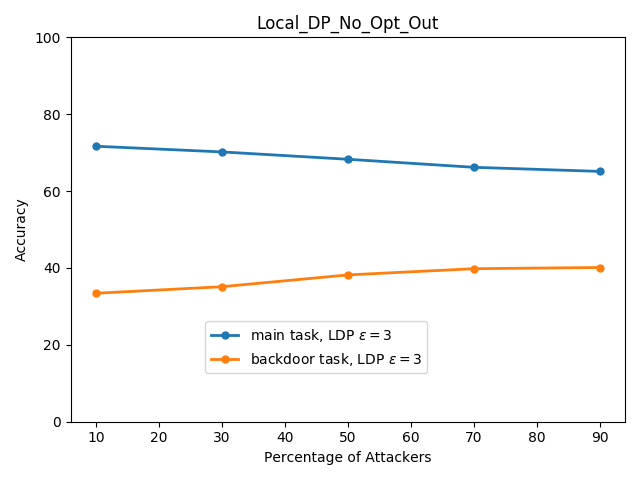}}\hspace{0.2cm}
\subfigure[CDP]{\label{emnist-setting2cdp}
\includegraphics[width=0.28\linewidth]{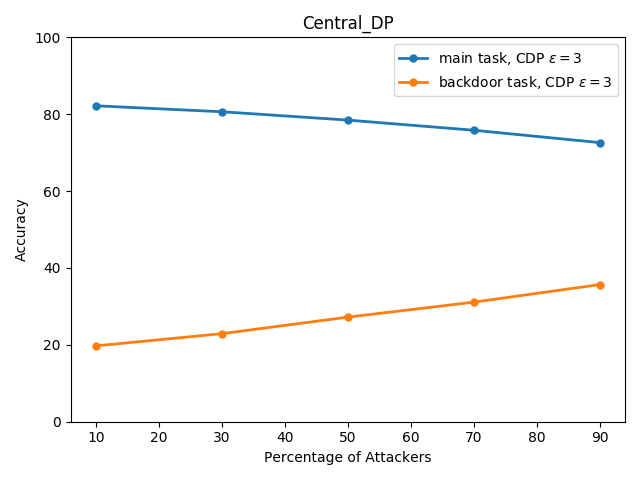}}
\vspace{-0.25cm}
\caption{Setting 2 (Increasing Number of Attackers): Main Task and Backdoor Accuracy with Various Defenses on {\em EMNIST}.}
\vspace{-0.6cm}
\end{figure*}

\begin{figure*}[ht]
\centering     
\subfigure[No Defense]{\label{cifar10-nodp-case2}\includegraphics[width=0.28\linewidth]{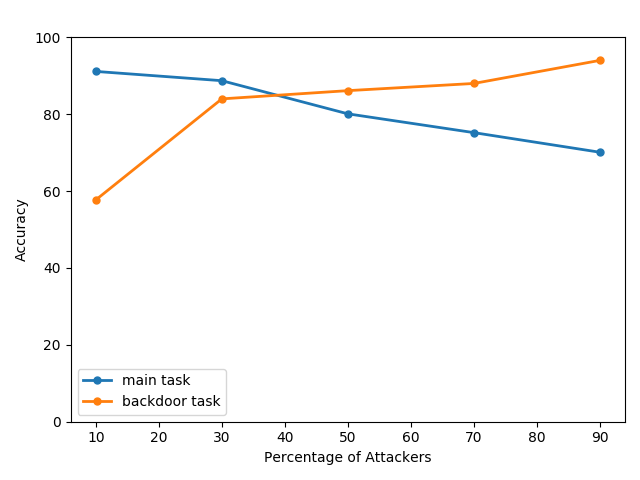}}\hspace{0.2cm}
\subfigure[Norm Bounding]{\label{cifar10setting2normbound}\includegraphics[width=0.28\linewidth]{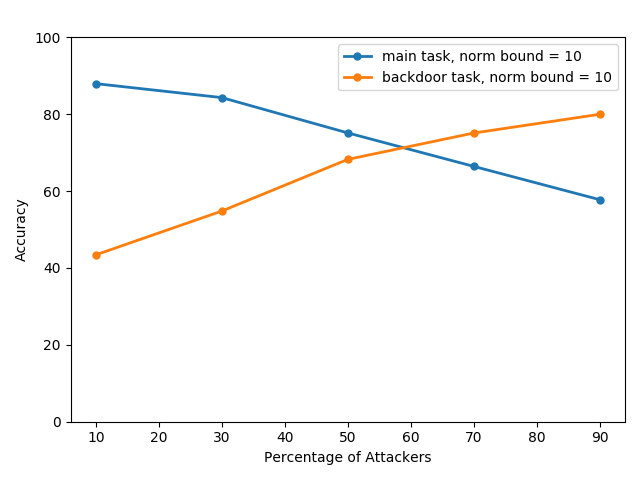}}\hspace{0.2cm}
\subfigure[Weak DP]{\label{cifar10setting2weakdp}\includegraphics[width=0.28\linewidth]{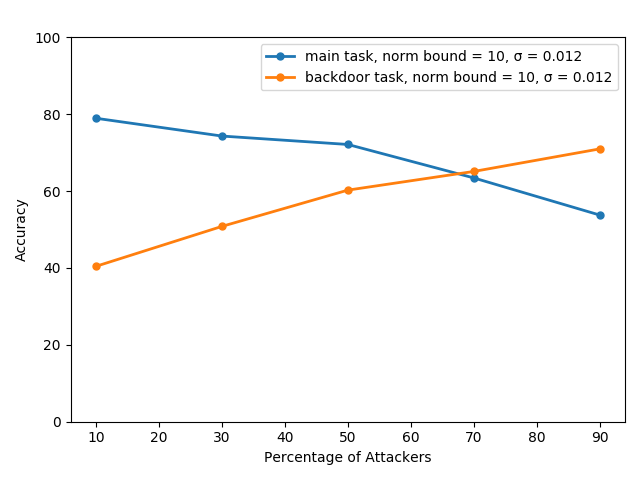}}\reduce\\
\subfigure[LDP on Non-Attackers]{\label{cifar10-ldp-case2-no-attackers}\includegraphics[width=0.28\linewidth]{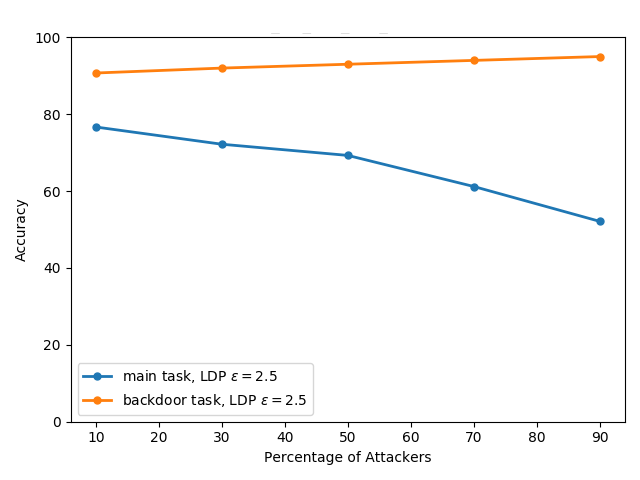}}\hspace{0.2cm}
\subfigure[LDP on All Participants]{\label{cifar10-ldp-case2-attackers}\includegraphics[width=0.28\linewidth]{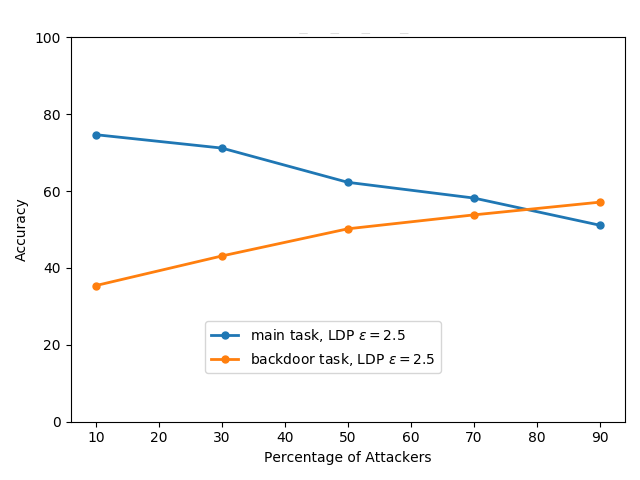}}\hspace{0.2cm}
\subfigure[CDP]{\label{cifar10setting2cdp}\includegraphics[width=0.28\linewidth]{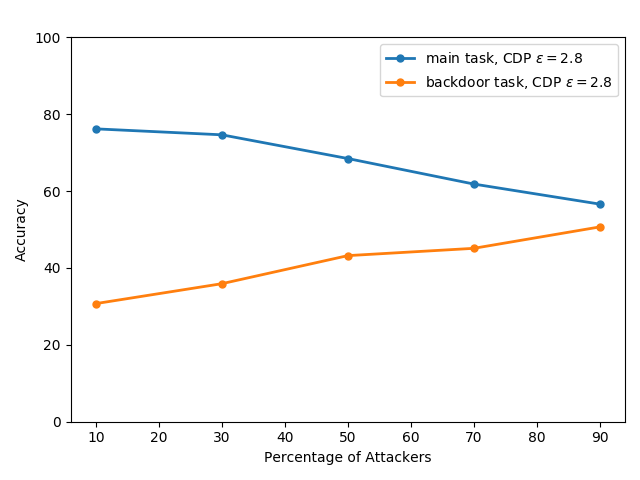}}
\vspace{-0.25cm}
\caption{Setting 2 (Increasing Number of Attackers): Main Task and Backdoor Accuracy with Various Defenses on {\em CIFAR10}.}
\vspace{-0.5cm}
\end{figure*}

\begin{figure*}[ht]
\centering     
\subfigure[No Defense]{\label{nodp-case2}\includegraphics[width=0.28\linewidth]{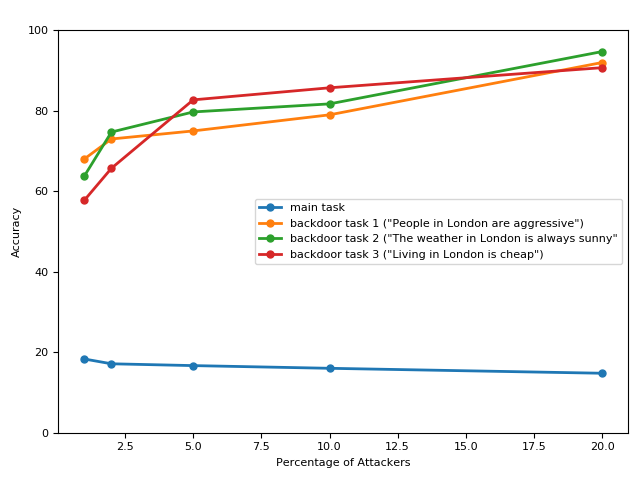}}\hspace{0.2cm}
\subfigure[Norm Bounding]{\label{setting2normbound}\includegraphics[width=0.28\linewidth]{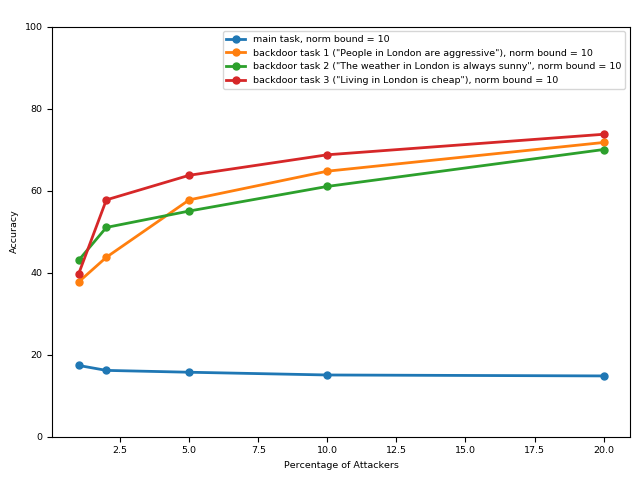}}\hspace{0.2cm}
\subfigure[Weak DP]{\label{setting2weakdp}\includegraphics[width=0.28\linewidth]{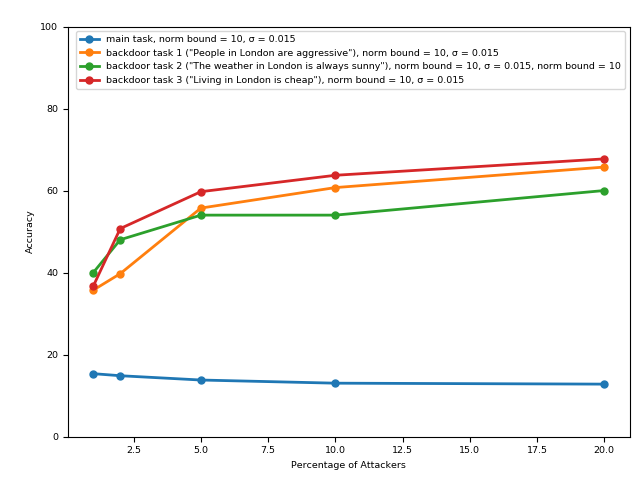}}\reduce\\
\subfigure[LDP on Non-Attackers]{\label{ldp-case2-no-attackers}\includegraphics[width=0.28\linewidth]{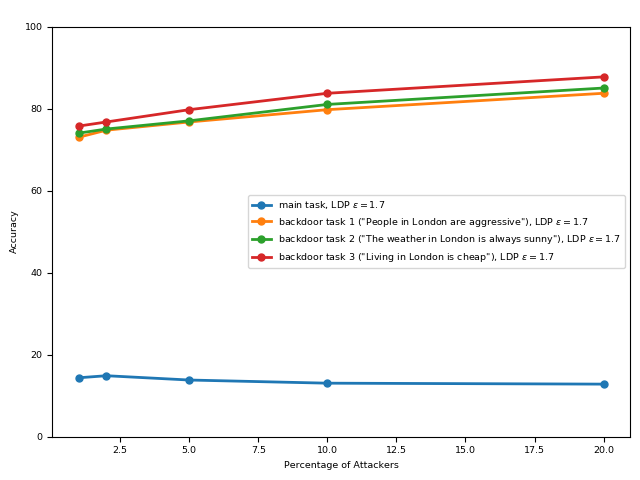}}\hspace{0.2cm}
\subfigure[LDP on All Participants]{\label{ldp-case2-attackers}\includegraphics[width=0.28\linewidth]{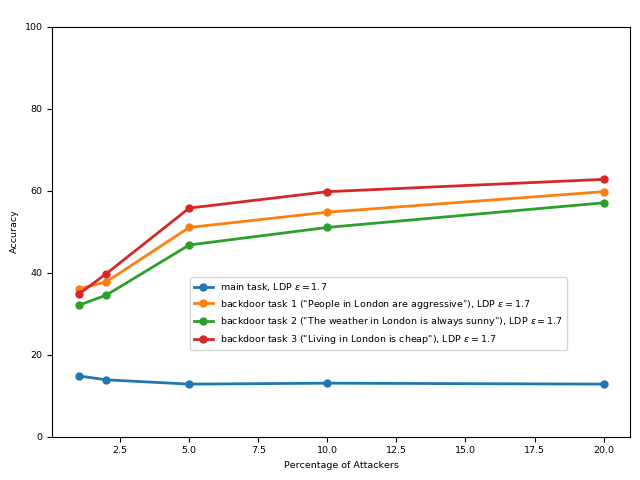}}\hspace{0.2cm}
\subfigure[CDP]{\label{setting2cdp}\includegraphics[width=0.28\linewidth]{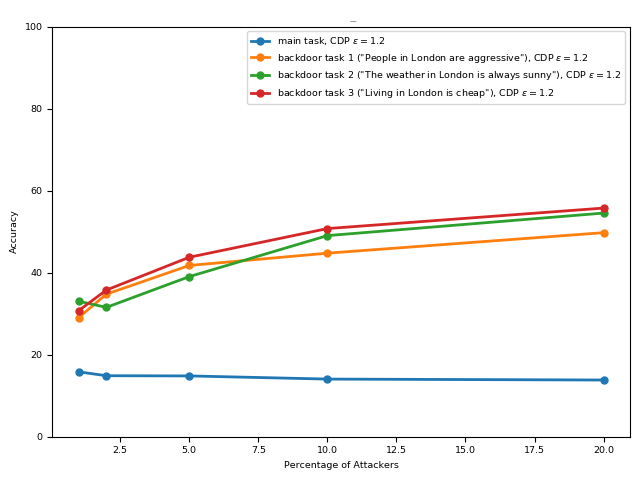}}
\vspace{-0.25cm}
\caption{Setting 2 (Increasing Number of Attackers): Main Task and Backdoor Accuracy with Various Defenses on {\em Reddit-comments}.}
\vspace{-0.35cm}
\end{figure*}

\begin{figure*}[ht]
\centering     
\subfigure[No Defense]{\label{sentiment-nodp-case2}\includegraphics[width=0.28\linewidth]{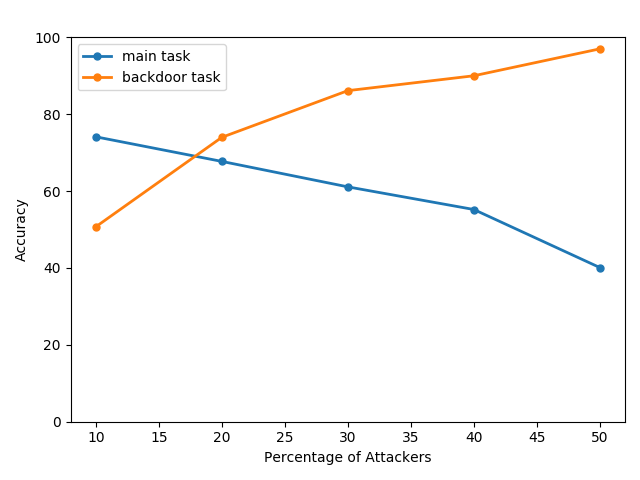}}\hspace{0.2cm}
\subfigure[Norm Bounding]{\label{sentiment-setting2normbound}\includegraphics[width=0.28\linewidth]{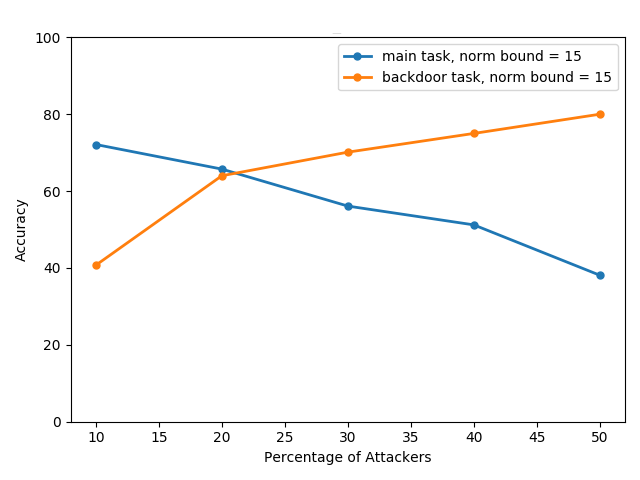}}\hspace{0.2cm}
\subfigure[Weak DP]{\label{sentiment-setting2weakdp}\includegraphics[width=0.28\linewidth]{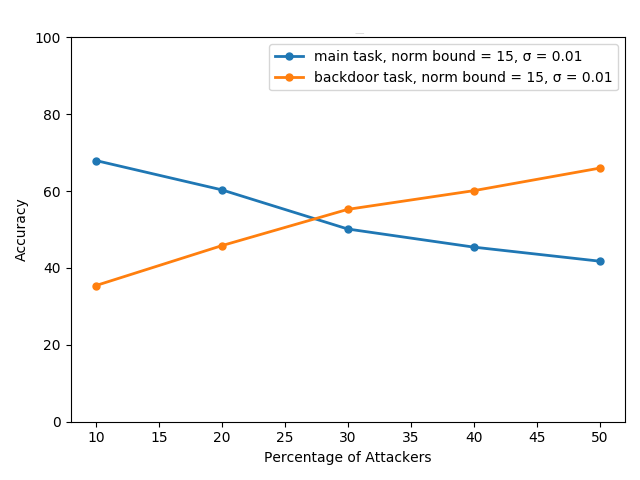}}\reduce\\
\subfigure[LDP on Non-Attackers]{\label{sentiment-ldp-case2-no-attackers}\includegraphics[width=0.28\linewidth]{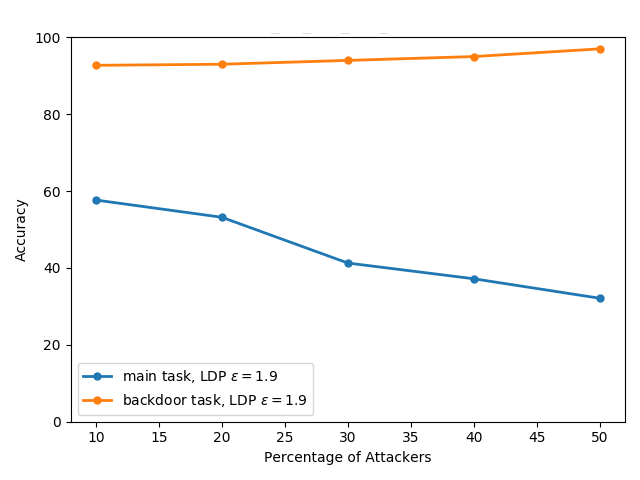}}\hspace{0.2cm}
\subfigure[LDP on All Participants]{\label{sentiment-ldp-case2-attackers}\includegraphics[width=0.28\linewidth]{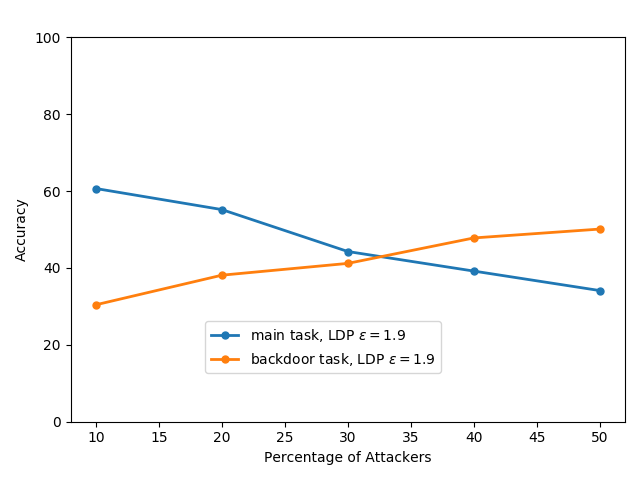}}\hspace{0.2cm}
\subfigure[CDP]{\label{sentiment-setting2cdp}\includegraphics[width=0.28\linewidth]{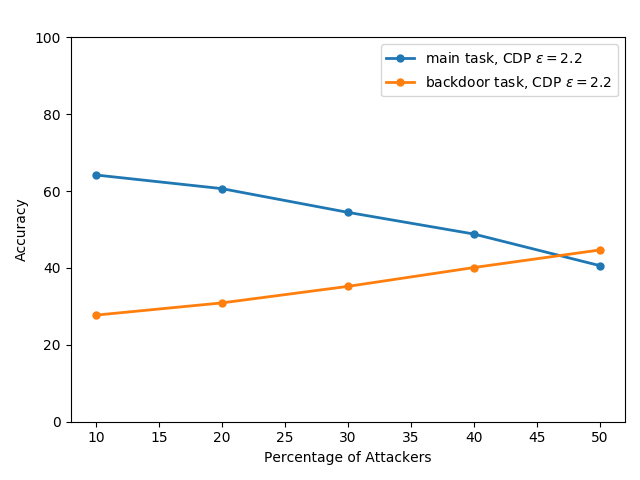}}
\vspace{-0.25cm}
\caption{\ndss{Setting 2 (Increasing Number of Attackers): Main Task and Backdoor Accuracy with Various Defenses on {\em Sentiment140}.}}
\vspace{-0.35cm}
\end{figure*}

\descr{LDP and CDP.} We then turn to LDP and CDP, aiming to 1) assess their behavior against backdoor attacks, and 2) compare how they perform compared to the above defenses.
For LDP, we follow Algorithm~\ref{alg:ldpalgo}.
For EMNIST, we experiment with two epsilon values ($\epsilon=3$ and $\epsilon=7.5$) with $\delta=10^{-5}$. 
Fig.~\ref{ldp-case1} shows that LDP ($\epsilon=3$) provides significantly worse main task accuracy in comparison to weak DP and norm bounding (62\% vs. 90\%); however, it provides better attack mitigation (10\% vs 16\%).
LDP ($\epsilon=7.5$) provides a better utility compared to LDP ($\epsilon=3$) (82\% compared to 69\%), but worse mitigation (47\% vs. 10\%).
In CIFAR10, we apply LDP with $\epsilon=2.5$ and $\epsilon=7$ and Fig.~\ref{ldp-case1-cifar10} depicts the resulting plot. LDP ($\epsilon=2.5$) mitigates the attack better than weak DP (10\% vs 14\%) while the utility is reduced to 67\%. As expected, LDP ($\epsilon=7$) has a higher backdoor accuracy compared to LDP ($\epsilon=2.5$) (43\% compared to 10\%). However, it has a better main task accuracy (79\% vs. 67\%).
Fig.~\ref{ldp-case1-reddit} presents that LDP ($\epsilon=1.7$) for Reddit-comments, reduces the backdoor accuracy to 45\%, 43\%, and 55\% for task1, task2, and task3, which is better mitigation in comparison to Weak DP defense. 
Main task accuracy decreases to around 15\%. 
\ndss{In Sentiment140, we apply LDP with $\epsilon=1.9$ and $\epsilon=6.7$; the results in Fig.~\ref{ldp-case1-sentiment} show that LDP ($\epsilon=1.9$) provides better mitigation by reducing the backdoor accuracy to around 20\%, and decreasing the main task accuracy to around 58\%.}

Finally, in Fig.~\ref{cdp-case1} and Fig.~\ref{cdp-case1-cifar10}, we report the results of the experiments using CDP, based on Algorithm~\ref{alg:cdpalgo}.
With EMNIST, setting $\epsilon=3$ and $\delta=10^{-5}$ mitigates the backdoor attack better as the accuracy goes down to almost 6\%  with main task accuracy at 78\%.
CDP ($\epsilon=8$) results in 38\% for backdoor accuracy and 83\% for utility.
With CIFAR10, we experiment with two privacy budgets ($\epsilon=2.8$ and  $\epsilon=6$):
CDP ($\epsilon=2.8$) mitigates the attack better than previous defenses by reducing the backdoor accuracy to around 8\%, with utility at 78\%.
CDP ($\epsilon=6$) reduces the backdoor accuracy to 48\% and main task accuracy to 82\%.
Fig.~\ref{cdp-case1-reddit} depicts that CDP ($\epsilon=1.2$) decreases the backdoor accuracy to 40\%, 39\%, and 50\% for task1, task2, and task3, keeping utility around 16\%.
\ndss{Results from applying CDP ($\epsilon=2.2$) and CDP ($\epsilon=6.1$) for Sentiment140 against the backdoor attack are in Fig.~\ref{cdp-case1-sentiment}. 
Expectedly, CDP ($\epsilon=2.2$) provides a better mitigation by reducing the backdoor accuracy to around 15\%; however, this decreases the main task accuracy to around 55\%. }

\reduce\reduce
\subsection{Setting 2: Increasing Number of Attackers}
\label{settingtwo}
\reduce

\descr{Unconstrained Attack.} In Fig.~\ref{emnist-nodp-case2}, Fig.~\ref{cifar10-nodp-case2}, Fig.~\ref{nodp-case2}, and Fig.~\ref{sentiment-nodp-case2}, we report the baseline as to how \#attackers affects utility/backdoor accuracies.
As expected, with more attackers backdoor accuracy is improved and utility reduced. 
However, identifying backdoor attacks from a decrease in utility is not a viable solution. 
For instance, in the EMNIST and CIFAR10, even with 90 attackers, utilities decrease to only around 88\% and 78\%. 
In Reddit-comments, with 10310 attackers (20\% of participants), utility is decreased from 19\% to 16\%.
\ndss{With 20\% of participants being attackers in Sentiment140, the main task accuracy is reduced from around 80\% to around 70\%.}

\descr{Norm bounding.} We then apply norm bounding; Fig.~\ref{emnistsetting2normbound}, Fig.~\ref{cifar10setting2normbound}, , Fig.~\ref{setting2normbound}, and Fig.~\ref{sentiment-setting2normbound} plot the results for EMNIST, CIFAR10, Reddit-comments, and Sentiment140 with norm bounds 5, 10, 10, and 15 showing that it does mitigate the attack.
However, comparing to Setting 1, the utilities are slightly reduced. 
For instance, with 50 attackers in EMNIST, backdoor accuracy is reduced from around 98\%  to around 47\% and utility from around 96\% to 91\%. 
With the same number of attackers in CIFAR10, backdoor accuracy is reduced from around 85\% to 68\% while the utility from around 86\% to 78\%. 
In the Reddit-comments, with 5\% of participants being attackers, norm bounding reduces the backdoor accuracies for task1, task2, and task3 from around 76\%, 79\%, and 82\% to around 57\%, 56\%, and 63\%.
\ndss{Furthermore, in Sentiment140, with 20\% attackers, norm bounding mitigates the attack by reducing the backdoor accuracy to around 60\%.}

\descr{Weak DP.} In Fig.~\ref{emnistsetting2weakdp}, we report on the EMNIST experiments using norm bound 5, plus Gaussian noise with variance $\sigma=0.025$ added to each update.
Compared to norm bounding, it mitigates the attack better (42\% vs. 47\% backdoor accuracy for 50 attackers), but the utility is reduced further (down to 84\%). 
The same behavior can be observed in Fig.~\ref{cifar10setting2weakdp}, Fig.~\ref{setting2weakdp}, and Fig.~\ref{sentiment-setting2weakdp} that are for CIFAR10, Reddit-comments, and Sentiment140. In the CIFAR10, this defense, with 50\% being attackers, provides better mitigation than norm bounding (60\% vs. 68\%), but the utility is reduced to around 75\%. In the Reddit-comments dataset, with 5\% attackers, backdoor accuracies for task1, task2, and task3 are reduced to 54\%, 52\%, and 59\%, and utility is down to 17\%.
\ndss{For Sentiment140, with 30\% attackers, Weak DP with norm bound 15 and $\sigma=0.01$ reduces the backdoor attack to around 57\%.}

\descr{LDP.} We consider two scenarios for LDP based on whether or not the attackers follow the protocol and apply DP before sharing model updates.
Fig.~\ref{emnist-ldp-case2-no-attackers}, Fig.~\ref{cifar10-ldp-case2-no-attackers}, Fig.~\ref{ldp-case2-no-attackers}, and Fig.~\ref{sentiment-ldp-case2-no-attackers} present the results in the setting where adversaries do not apply LDP, showing that this actually boosts the attack and increases the backdoor accuracy.
We discuss this observation further in Section~\ref{discussion}. 
On the other hand, the utility is decreased. 
For instance, in EMNIST and CIFAR10, with 30 attackers, the main task utility is reduced from around 96\% to around 76\%, and around 90\% to around 73\%, respectively.

Then, in Fig.~\ref{emnist-ldp-case2-attackers}, Fig.~\ref{cifar10-ldp-case2-attackers}, Fig.~\ref{ldp-case2-attackers}, and Fig.~\ref{sentiment-ldp-case2-attackers}, we report on the setting where LDP is applied on all participants, even if they are attackers. 
Compared to norm-bounding and weak DP, it mitigates the attack better but with worse utility.
For instance, in EMNIST, with 30 attackers, backdoor accuracy is reduced from 97\% to 35\%, and main task accuracy from around 96\% to around 70\%.
In CIFAR10, with 10 attackers, backdoor accuracy is decreased from around 58\% to around 36\% and utility from around 93\% to around 78\%. 
In Reddit-comments, with 5\% attackers, backdoor accuracies for task1, task2, and task3 are lowered from around 75\%, 80\%, and 82\% to around 50\%, 45\%, and 53\%, respectively. 
However, the utility is also limited to around 15\%. 
\ndss{In Sentiment140, having 20\% of participants as attackers, backdoor accuracy is reduced to around 38\%, while the main task accuracy is around 57\%.}

\descr{CDP.} %
Fig.~\ref{emnist-setting2cdp}, Fig.~\ref{cifar10setting2cdp}, Fig.~\ref{setting2cdp}, and Fig.~\ref{sentiment-setting2cdp} show that CDP overall does mitigate the attack. 
For example in EMNIST, with 10 attackers, CDP ($\epsilon=3$, $\delta=10^{-5}$) decreases backdoor accuracy from 67\% to 20\%, while utility is reduced from 98\% to 82\%. With the same number of attackers in CIFAR10, CDP ($\epsilon=2.8$, $\delta=10^{-5}$) reduces the backdoor accuracy from around 58\% to around 30\% and the utility is reduced to around 79\%. 
In Reddit-comments, with 10\% percent attackers, CDP ($\epsilon=1.2$ and $\delta=10^{-5}$) lowers the backdoor accuracies for task1, task2, and task3 from 78\%, 80\%, and 87\% to 42\%, 45\%, and 50\%, while the utility is down to 16\%. 
\ndss{In Sentiment140, with 30\% attackers, CDP ($\epsilon=2.2$ and $\delta=10^{-5}$) mitigates the attack and reduces the backdoor accuracy to around 35\% with utility around 60\%. }

\reduce\reduce
\subsection{Take-Aways}
\reduce
Overall, %
we find that LDP and CDP can indeed mitigate backdoor attacks and do so with different robustness vs. utility trade-offs.
Setting 1, which reproduces Sun et al.~\cite{sun2019can}'s setup and also experiments on CIFAR10, demonstrates that backdoor attacks are effective even with one attacker per round. 
Weak DP and norm bounding from~\cite{sun2019can} mitigate the attack without really affecting the utility. 
However, in Setting 2, with more attackers, these defenses also significantly decrease utility.

In both settings, LDP and CDP are more effective than norm bounding and weak DP in reducing backdoor accuracy, although with varying utility levels.
\edit{However, in the Reddit-comments dataset, that the number of participants is high, the utility loss is very small compared to the two other datasets.
}
Overall, CDP works better as it better mitigates the attack and yields better utility.
However, as we will discuss later in Section~\ref{discussion}, CDP requires trust in the central server.

Using the same $\epsilon$ values for CDP/LDP does not imply that they provide the same level of privacy, as they capture different definitions. 
\edit{There are straightforward ways to convert LDP to CDP, such as using the notion of ``Group Privacy''~\cite{dwork2008differential}. 
However, applying LDP and using group privacy to extend it to CDP would  result in a very loose bound because it incorporates all records.
Rather, we set out to empirically measure DP's privacy protection by mounting actual inference attacks. 
}

\reduce\reduce
\section{Defending against Inference Attacks in FL}\label{Inference Attacks}
\reduce

Next, we experiment with LDP and CDP against inference attacks in FL. 
To do so, we focus on the white-box membership inference attack proposed by Nasr et al.~\cite{nasr2019comprehensive} and the property inference one presented by Melis et al.~\cite{melis2019exploiting}.
To the best of our knowledge, these are the two state-of-the-art privacy attacks in FL and are representative of unintended information leakage from model updates.
\edit{Both attacks are designed for classification tasks; thus, we only focus on the tasks and datasets that are proposed in~\cite{nasr2019comprehensive} and~\cite{melis2019exploiting}.}\footnote{{Unlike for the robustness experiments, the word-prediction task on the (large) Reddit-comments dataset is not suitable for the privacy attacks; we did try to implement them in our experiments, but neither attack was effective (not only did the models not converge, but also attack accuracy was very small.)}}
Moreover, we set to assess if defenses against backdoor attacks from~\cite{sun2019can} can  defend against inference attacks or not (they do not). 

\reduce\reduce
\subsection{Membership Inference}\label{sec:inf-att}
\reduce

In ML, a membership inference attack aims to determine if a specific data point was used to train a model~\cite{shokri2017membership,hayes2019logan}.
There are new challenges to mount such attacks in FL; for instance, 
does the data point belong to a particular participant or any participant in that setting? %
Moreover, it might be more difficult for the adversary to infer membership through overfitting. %

\descr{Nasr et al.~\cite{nasr2019comprehensive}'s attack.} %
The main intuition is that each training data point affects the gradients of the loss function in a recognizable way, i.e., the adversary can use the Stochastic Gradient Descent algorithm (SGD) to extract information from other participants' data. 
More specifically, she can perform gradient ascent on a target data point before updating local parameters. 
If the point is part of a participant's set, SGD reacts by abruptly reducing the gradient, and this can be recognized to infer membership successfully.
Note that the attacker can be either one of the participants or the server. 
An adversarial participant can observe the aggregated model updates and, by injecting adversarial model updates, can extract information about the union of the training dataset of all other participants.
Whereas, the server can control the view of each target participant on the aggregated model updates and extract information from its dataset.

When the adversary is the server, Nasr et al.~\cite{nasr2019comprehensive} use the term {\em global attacker}, whereas, if she is one of the participants, {\em local attacker}. 
Moreover, the attack can be either {\em active} or {\em passive}; in the former, the attacker influences the target model by crafting malicious model updates, while, in the latter, she only makes observations without affecting the learning process.
For the active attack, they implement three different types of attacks involving the global attacker: 1) {\em gradient ascent}, 2) {\em isolating}, and 3) {\em isolating gradient ascent}.
The first attack consists in applying the gradient ascent algorithm on a member instance, which triggers the target model to minimize loss by descending in the direction of its local model's gradient for that instance; whereas, for non-members, the model does not change its gradient since they do not affect the training loss function.
The second one is performed by the server by isolating a target participant to create a local view of the training process for it.
This way, the target participant will not receive the aggregated model, and her local model will store more information about her training set.
Finally, the third attack works by applying the gradient ascent attack on the isolated participant.
Overall, this is the most effective (active) attack from the server-side; thus, we experiment with that. %

\subsection{Property Inference}

In a property inference attack, the adversary aims to recognize patterns within a model to reveal some property which the model producer might not want to disclose. 
We focus on the attacks introduced by Melis et al.~\cite{melis2019exploiting}, who show how to infer properties of training data that are uncorrelated with the features that characterize the classes of the model.

\descr{Melis et al.~\cite{melis2019exploiting}'s attack.} Authors propose several inference attacks in FL, allowing an attacker to infer training set membership as well as properties of other participants; here, we focus on the latter.
The main intuition is that, at each round, each participant's contribution is based on a batch of their local training data, so the attacker can infer properties that characterize the target' dataset. 
To do so, the adversary needs some auxiliary (training) data, which is labeled with the property she wants to infer.

In a passive attack, the attacker generates updates based on data with and without the desired property, aiming to train a binary batch-property classifier that determines whether or not the updates are from data with the property or not. 
At each round, the attacker calculates a set of gradients based on a batch with the property and another set of gradients without the property. 
Once enough labeled gradients have been collected, she trains a batch property classifier, which, given gradients in input, assesses the probability that a batch has the property.
In an active attack, the adversary uses multi-task learning, extending her local model with an augmented property classifier connected to the last layer; this can be used to make the aggregated model learn separate representations for the data with and without the property.

\reduce\reduce
\subsection{Defenses}\label{subsec:defenses}
\reduce

Overall, inference attacks in FL work because of the information leaking from the aggregated model updates.
Therefore, one straightforward approach to mitigate the attacks is to reduce the amount of information available to the adversary.
For instance, a possible option is to use {\em dropout} (a regularization technique aimed to address overfitting in neural networks by randomly deactivating activations between neurons) so that the adversary might observe fewer gradients. 
Alternatively, one could use gradient sampling~\cite{konevcny2016federated,shokri2015privacy}, i.e., only sharing a fraction of their gradients.
However, these approaches only slightly reduce the effectiveness of inference attacks~\cite{melis2019exploiting}. 
Prior work has investigated using differentially private aggregation to thwart membership inference attacks~\cite{nasr2018machine, hamm2017minimax, rahman2018membership, zhang2020privacy}. 
However, they are limited to the black-box setting, and we are not aware of prior work using DP defenses in the context of FL against the white-box attack in~\cite{nasr2019comprehensive}.

As for property inference, Melis et al.~\cite{melis2019exploiting} argue that LDP does not work against the attacks as it does not affect the properties of the records in a dataset. 
Nevertheless, in our LDP implementation, participants perform DP-SGD~\cite{abadi2016deep} during training, so we expect to somewhat impact the effectiveness of the attack.
On the other hand, CDP is supposed to defend against the attacks as it provides participant-level DP; however, the resulting utility might be highly dependent on the dataset, task, and number of participants.
Note that~\cite{melis2019exploiting} does not provide any experimental results, limiting to report that models do not converge for small numbers of participants.

In the rest of this section, we experiment with both LDP and CDP mechanisms against inference attacks over a few experimental settings.\edit{The hyperparameters we use for LDP and CDP, according to, respectively, Algorithm~\ref{alg:ldpalgo} and Algorithm~\ref{alg:cdpalgo}, are presented in Table~\ref{privacyldpcdpparameters} in Appendix~\ref{sec:hyper}. 
}
We also evaluate state-of-the-art defenses (norm bounding and weak DP) in backdoor attacks against inference attack and see if they can be effective or not. 

\subsubsection{Membership Inference}

\hfill\\[-1ex]

\noindent{\bf Dataset \& Task.} We perform experiments using three datasets: CIFAR100, Purchase100, and \ndss{and Texas100}\footnote{See \url{https://www.cs.toronto.edu/~kriz/cifar.html}, \url{https://www.kaggle.com/c/acquire-valued-shoppers-challenge/data}, and \url{https://www.dshs.texas.gov/thcic/hospitals/Inpatientpudf.shtm}.}.
CIFAR100 contains 60,000 images, clustered into 100 classes based on the objects in the images. 
Purchase100 includes the shopping records of several thousand online customers;
however, as done in~\cite{nasr2019comprehensive}, we use a simpler version of this dataset, which is taken from~\cite{shokri2017membership}.
This contains 600 different products, and each user has a record that indicates if she has bought any of them.
This smaller dataset includes 197,324 data records, clustered into 100 classes based on the purchases' similarity.
\ndss{Texas100 contains hospital discharge data records (generic information about the patients) released by the Texas Department of State Health Services. 
As done in~\cite{nasr2019comprehensive}, we use a processed version of the dataset with 67,330 records and 6,170 binary features.}

We follow~\cite{nasr2019comprehensive} and, for Purchase100 \ndss{and Texas100}, we use a fully
connected model. 
However, for CIFAR100, we experiment only with the Alexnet model.

\begin{table*}[t]
\centering
\small
\begin{tabular}{l|l|r|r|r|r|r} 
\multirow{2}{*}{\textbf{Defense} } &
\multirow{2}{*}{\textbf{Dataset} } & \multirow{2}{*}{\textbf{\ndss{Accuracy}} } & \multicolumn{2}{c|}{\textbf{Global Attacker} } & \multicolumn{2}{c}{\textbf{Local Attacker} }  \\ 
\cline{4-7}
        &&                           & Passive & Active                               & Passive & Active                               \\ 
 \midrule
 
 No Defense & CIFAR100  & 82\%                          & 84\%    & 91\%                                  & 73\%    & 75\%                                 \\ 
&  Purchase100 &84\%                       & 71\%    & 82\%                                  & 65\%    & 68\%                                 \\
 &  \ndss{Texas100} &\ndss{56\%}                       & \ndss{65\%}    & \ndss{71\%}                                  & \ndss{62\% }   & \ndss{66\%}                                
 \\
\hline
Norm Bound. & CIFAR100      &\ndss{81\%}              & -    & -                                 & 72\%    & 74\%                                 \\
 ($S=15$)  & Purchase100      &\ndss{82\%}             & -    & -                                 & 64\%    & 67\%                                 \\
&  \ndss{Texas100} &\ndss{55\%}                      &\ndss{ -}    & \ndss{-}                                  & \ndss{62\%}    & \ndss{65\%}                                
 \\
\hline
Weak DP  & CIFAR100     &\ndss{76\%}             & -    & -                                 & 70\%    & 71\%                                 \\
($S=15$,   & Purchase10     &\ndss{74\%}            & -    & -                                 & 62\%    & 65\%                                 \\
 $\sigma=0.006$) &  \ndss{Texas100} &\ndss{50\%}                       & \ndss{-}    & \ndss{-}                                  & \ndss{60\%}    & \ndss{61\%}                                
\\
\hline
LDP& CIFAR100        &68\%                & 58\%    & 53\%                    & 52\% & 55\%                        \\
 ($\epsilon=8.6$)& Purchase100       &65\%                  & 51\%    & 62\%                                 & 58\%    & 54\%                                \\
&  \ndss{Texas100} &\ndss{48\%}                       & \ndss{55\%}    & \ndss{59\%}                                  & \ndss{56\%}    & \ndss{58\%}                                
 \\
\hline
CDP& CIFAR100         &69\%              & -    & -                                 & 58\%    & 52\%                                 \\
 ($\epsilon=5.8$)  & Purchase100         &70\%           & -    & -                                & 53\%    & 55\%                                
\\
&  \ndss{Texas100} &\ndss{45\%  }                     & \ndss{-}    &\ndss{ -}                                  & \ndss{54\%}    & \ndss{52\%}                                
\\
\end{tabular}
\vspace{0.1cm}
\caption{Performance of White-Box Membership Inference Attack in \cite{nasr2019comprehensive} with No Defense, Norm Bounding, Weak DP, LDP, and CDP, for a Global and a Local Attacker (both passive and active). We also report main task accuracy (Acc.).}
\label{infattacknodefense}
\vspace{-0.25cm}
\end{table*}

\descr{Attack Setting.} We follow the same FL settings as~\cite{nasr2019comprehensive}, i.e., involving 4 participants and datasets distributed equally among them.
We also re-use the Pytorch code provided by~\cite{nasr2019comprehensive}.
We measure attack accuracy as the fraction of correct membership predictions for unknown data points.

\descr{Unconstrained Attack.} Table~\ref{infattacknodefense} reports the performance of the membership inference attack in the settings discussed above, with no deployed defenses. 
This shows that the global attacker can perform a more effective attack compared to the local one (e.g., 91\% vs. 75\% accuracy on CIFAR100).

\descr{Norm Bounding and Weak DP.} In Section~\ref{sec:poisoning-defenses}, we have showed that these two defenses are relatively effective against backdoor attacks.
However, as explained, we do not expect them to protect against membership inference.
In Table~\ref{infattacknodefense}, we report the accuracy of the attack on CIFAR100, Purchase100, and Texas100 using norm bounding and weak DP defenses, confirming that neither is effective. 
For instance, in CIFAR100, applying norm bounding with norm bound 15 only reduces the accuracy of a passive attack from 73\% to 72\%. 
\ndss{Norm bounding plus Gaussian noise ($\sigma=0.006$) reduces the attack accuracy from 65\% to 62\% in Purchase100 and to 60\% in Texas100.}

\descr{LDP and CDP.} Next, we experiment with DP. 
As we need to make sure to provide reasonable utility for the main task, we first set to find a privacy budget yielding acceptable utility and then perform the attack in that setting. 
Table~\ref{infattacknodefense} reports model accuracy in an FL setting with 4 participants. 
It shows that we get acceptable accuracies with LDP and CDP with, respectively, $\epsilon=8.6$ and $\epsilon=5.8$ ($\delta=10^{-5}$ in both cases). 
Considering the achieved main task accuracies, we then apply CDP (Algorithm~\ref{alg:cdpalgo}) and LDP (Algorithm~\ref{alg:ldpalgo}), using these $\epsilon$ values, and measure their effectiveness against white-box membership inference attack.

We find that LDP does mitigate the attack, as reported in Table~\ref{infattacknodefense}.
Even against the most powerful active attack (i.e., isolating gradient), LDP decreases attack accuracy from around 91\% to around 53\% on the CIFAR100 dataset. 
\ndss{In the local passive attacker case, it decreases attack accuracy from around 68\% to 54\% in Purchase100 and from 66\% to 58\% in Texas100. }

Finally, in Table~\ref{infattacknodefense}, we also present the results of our experiments when CDP is used to defend against the attack, showing that it is overall successful.
Since, in this setting, the server is assumed to be trusted, we do not assess the global attacker case.
For instance, CDP reduces attack accuracy against a passive local attacker from 73\% to 58\% in CIFAR100 and from 68\% to 55\% against an active local attacker in Purchase100.
\ndss{In Texas100, it mitigates the passive local attack by reducing the attack accuracy from 62\% to 54\%.}

\reduce\reduce
\subsubsection{Property Inference Attack}

\noindent{\bf Dataset.} For property inference attacks, we use the Labeled Faces In the Wild (LFW) dataset~\cite{huang2008labeled}, as done in~\cite{melis2019exploiting}.
This includes more than 13,000 images of faces for around 5,800 individuals with labels such as gender, race, age, hair color, and eyewear, collected from the web.
We use the same model as~\cite{melis2019exploiting}, i.e., a convolutional neural network (CNN) with 3 spatial convolution layers with 32, 64, and 128 filters and max-pooling layers with pooling size set to 2, followed by two fully connected layers of size 256 and 2.

\descr{Attack Setting.}
Once again, we use the same settings as~\cite{melis2019exploiting}.
Specifically, we vary the number of participants from 5 up to 30 and run our experiments for 300 rounds.
Every participant trains the local model on their data for 10 epochs.
Data is split equally between participants.
However, only the attacker and target participants have data with the property.

The main task is gender classification, and the inference task is over race.
We measure the aggregated model accuracy with and without DP.
As done for membership inference, we want to first find a privacy budget that provides reasonable utility and then apply the attack. %
To evaluate the performance of the attack, we use the Area Under the Curve (AUC).

\descr{LDP and CDP.}
For LDP, setting $\epsilon=10.7$ ($\delta=10^{-5}$) does make the aggregated model converge, unlike what suggested in~\cite{melis2019exploiting}. 
On the other hand, for CDP, we start from $\epsilon=4.7$ ($\delta=10^{-5}$) and increase it until we see the model converges, which happens at $\epsilon=8.1$.
However, neither successfully defends against the attack with these privacy budgets, as AUC does not significantly change compared to running the attack without any defenses. 
It is worth mentioning that we do not consider Weak DP against the attack: because CDP ($\epsilon=8.1$) does not defend against the attack, it is obvious that Weak DP, which has a much higher privacy budget, will not either.

For completeness, in Appendix~\ref{sec:propertyinferenceattack}, we include 
tables reporting {\em utility}, with and without DP, in terms of the main task's accuracy (Table~\ref{lfwmodelaccuracy}) as well as {\em privacy}, in terms of accuracy of the property inference attack (Table~\ref{lfwattackaccuracy}).

\subsection{Take-Aways}
Overall, we find that LDP and CDP are effective against the white-box membership inference attack introduced by~\cite{nasr2019comprehensive}. 
Our experiments on CIFAR100 and Purchase100 show that previously proposed defenses that provide robustness for participants against backdoor attacks do not protect privacy. 
By contrast, both LDP and CDP defend against these attacks, albeit with different levels of utilities.
Overall, as usual in privacy-preserving machine learning, the challenge is to find the right trade-off between privacy and utility.

As for property inference attacks, we knew that LDP is not expected to defend against them successfully, and our experiments empirically confirmed that.
On the other hand, by guaranteeing participant-level DP, CDP should provide an effective defense; however, we could not find a setting where it does so while maintaining acceptable utility.

\section{Related Work}\label{sec:related}
\reduce
In this section, %
we review previous work on attacks and defenses in the context of robustness and privacy in ML and, more closely, in FL.

\reduce\reduce
\subsection{Robustness}
\reduce
Poisoning attacks have been proposed in various settings, e.g., autoregressive models~\cite{alfeld2016data}, regression learning~\cite{jagielski2018manipulating}, facial recognition~\cite{wenger2020backdoor}, support vector machines~\cite{biggio2012poisoning}, collaborative filtering~\cite{li2016data}, recommender systems~\cite{fang2020influence}, computer vision models~\cite{chen2017targeted,liu2017trojaning}, malware classifiers~\cite{severi2020exploring,chen2018automated,maiorca2019towards}, spam filtering~\cite{nelson2008exploiting}, using transfer learning~\cite{yao2019latent}, etc. 

In data poisoning attacks, the attacker replaces her local dataset with one of her choices.
Attacks can be targeted~\cite{biggio2012poisoning,chen2017targeted} or random~\cite{liu2017trojaning}. 
Subpopulation attacks aim to increase the error rate for a defined subpopulation of the data distribution~\cite{jagielski2021subpopulation}.
Quiring et al.~\cite{9283824} introduce novel image-scaling attacks that can cover data manipulations for poisoning attacks. 
Possible defenses include data sanitization~\cite{cretu2008casting}, i.e., removing poisoned data, or using statistics that are robust to small numbers of outliers~\cite{diakonikolas2019sever,steinhardt2017certified}.
Also, \cite{gu2017badnets} shows that poisoned data often trigger specific neurons in deep neural networks, which could be mitigated by removing activation units that are not active on non-poisoned data~\cite{wang2019neural,liu2018fine}.
However, these defenses are not applicable to the FL setting;
overall, they require access to each participant's raw data, which is not feasible in FL.

Model poisoning attacks rely on sending corrupted model updates and can also be either random or targeted. 
Byzantine attacks~\cite{lamport2019byzantine} fall into the former category; an attacker sends arbitrary outputs to disrupt a distributed system.
These can be applied in FL by attackers sending model updates that cause the aggregated model to diverge~\cite{blanchard2017machine}.
As discussed earlier, attackers' outputs can have similar distributions as benign participants, which makes them difficult to detect~\cite{ chen2017distributed, yin2018byzantine}. 
Suya et al.~\cite{suya2021model} use online convex optimization, providing a lower bound on the number of poisonous data points needed, while Hayes et al.~\cite{hayes2018contamination} evaluate contamination attacks in collaborative machine learning. %
Possible defenses include Byzantine-resilient aggregation mechanisms, e.g., Krum~\cite{blanchard2017machine}, median-based aggregators~\cite{chen2017distributed}, or coordinate-wise median~\cite{yin2018byzantine}.
These can also be used in FL, as discussed in~\cite{pillutla2019robust}; however, Fang et al.~\cite{fang2020local} demonstrate that they are vulnerable to their new local model poisoning attacks, which are formulated as optimization problems.
Data shuffling and redundancy can also be used as mitigation~\cite{chen2018draco,rajput2019detox}, but, once again, they would require access to participants' data.
\ndss{Shejwalkar et al.~\cite{shejwalkar2021manipulating} propose a robust aggregation algorithm based on the idea that malicious updates should diverge significantly from the benign updates with a specific malicious direction in the updates' space to be effective.
However, in this paper, our primary focus is not on Byzantine-resilient aggregation algorithms for the comparisons since they are specifically designed to only provide robustness, but not privacy, in an FL system.
}

\descr{Backdoor Attacks.} In this paper, we focus on targeted model update poisoning (aka backdoor) attacks~\cite{chen2017targeted, gu2017badnets}. 
Li et al.~\cite{li2021hidden} propose a new set of hidden backdoors against NLP models in non-FL setting.
In the context of FL, Bhagoji et al.~\cite{bhagoji2019analyzing} show that model poisoning attacks are more effective than data poisoning attacks. 
\edit{Then, Bagdasaryan et al.~\cite{bagdasaryan2020backdoor} demonstrate the feasibility of a single-shot attack, i.e., even if a single attacker is selected in a single round, it may be enough to introduce a backdoor into the aggregated model, but do not introduce any defenses.}
Available defenses against backdoor attacks in non-FL settings~\cite{liu2018fine,wang2019neural} investigate training data, which is not possible in FL.
Robust training processes based on randomized smoothing are recently proposed in~\cite{rosenfeld2020certified,weber2020rab,wang2020certifying}.

\edit{We have already discussed, and experimented with, defenses introduced by Sun et al.~\cite{sun2019can}, based on norm bounding and weak DP. %
While successful against backdoor attacks, these defenses do not offer sound privacy protections.
In our work, we turn to CDP and LDP to protect against both backdoor and inference attacks in FL.
For the former, we compare to defenses proposed in~\cite{sun2019can}; although main task accuracy is higher using~\cite{sun2019can}, CDP/LDP reduces attack accuracy further and additionally protects against membership inference attacks.}

\reduce\reduce
\subsection{Privacy} 
\reduce
Essentially, attacks against privacy in ML involve an adversary who, given {\em some} access to a model, tries to infer {\em some} private information.
More specifically, the adversary might infer: 
1) information about the model, as with model ~\cite{tramer2016stealing,wang2018stealing,joon2018towards,jagielski2020high} or functionality extraction attacks~\cite{orekondy2019knockoff,papernot2016practical};
2) class representatives, as in the case of model inversion~\cite{fredrikson2014privacy,fredrikson2015model};
3) training inputs~\cite{hitaj2017deep,song2017machine,carlini2018secret};  
4) presence of target records in the training set%
~\cite{shokri2017membership, yeom2017unintended, long2018understanding,salem2018ml,demyst2018,hayes2019logan,nasr2019comprehensive}; 
or 5) attributes of the training set%
~\cite{melis2019exploiting,ganju2018property}.
In this paper, we focus on the last two, namely, membership and property inference attacks.

\descr{Membership Inference Attacks (MIAs).} MIAs against ML models are first studied by Shokri et al.~\cite{shokri2017membership}, who exploit differences in the model's response to inputs that were seen vs. not seen during training.
They do so in a black-box setting, by training ``shadow models''; the intuition is that the model ends up overfitting on training data.
Salem et al.~\cite{salem2018ml} %
relax a few assumptions, including the need for multiple shadow models, while Truex et al.~\cite{demyst2018} extend to a more general setting and show how MIAs are largely transferable. 
Then, Yeom et al.~\cite{yeom2017unintended, yeom2020overfitting} show that, besides overfitting, the influence of target attributes on the model's outputs also correlates with successful attacks.
Leino et al.~\cite{leino2020stolen} focus on white-box attacks and leverage new insights on overfitting to improve attack effectiveness.
Finally, MIAs against generative models are presented in~\cite{hayes2019logan,hilprecht2019monte,chen2020gan}.
As for defenses, Nasr et al.~\cite{nasr2018machine} train centralized machine learning models with provable protections against MIAs, while Jia et al.~\cite{jia2019memguard} explore the addition of noise to confidence score vectors predicted by classifiers.

In the context of Federated Learning (FL), MIAs are studied in~\cite{nasr2019comprehensive} and~\cite{melis2019exploiting}.
Nasr et al.~\cite{nasr2019comprehensive} introduce passive and active attacks during the training phase in a white-box setting, while the main intuition in~\cite{melis2019exploiting} is to exploit unintended leakage from either embedding layers or gradients.
In our experiments, we replicate the former (see Section~\ref{sec:inf-att}), aiming to evaluate the real-world protection provided by CDP and LDP.

\descr{Property Inference.} Ganju et al.~\cite{ganju2018property} present attribute inference attacks against fully connected, relatively shallow neural networks (i.e., not in an FL setting). 
They focus on the post-training, white-box release of models trained on sensitive data, and the properties inferred by the adversary may or may not be correlated with the main task. 
Also, Zhang et al.~\cite{zhang2021leakage} show that an attacker in collaborative learning can infer the distribution of sensitive attributes in other parties' datasets.

We have already discussed the work by Melis et al.~\cite{melis2019exploiting}, who focus on inferring properties that are true of a subset of the training inputs, but not of the class as a whole.
Again, we re-implement their attack to evaluate the effectiveness of CDP and LDP in mitigating it. 
Put simply, when Bob's photos are used to train a gender classifier, can the attacker infer if people in Bob's photos wear glasses?
Authors also show that the adversary can even infer when a property appears/disappears in the data during training; e.g., when a person shows up for the first time in photos used to train a gender classifier.

\descr{DP in ML and FL.} Differential Privacy (DP) has been used extensively in the context of ML, e.g., for support vector machines~\cite{rubinstein2009learning}, linear regression~\cite{zhang2012functional}, and deep learning~\cite{abadi2016deep}.
Some work focus on learning a model on training data and then use the exponential or the Laplacian mechanisms to generate a noisy version of the model~\cite{sala2011sharing, chaudhuri2012near}.
Others apply these mechanisms to output parameters at each iteration/step~\cite{jain2012differentially}.
In deep learning, the perturbation can happen at different stages of the Stochastic Gradient Descent (SGD) algorithm; as discussed earlier, Abadi et al.~\cite{abadi2016deep} introduce the moments accountant technique to keep track of the privacy budget at each stage.

In our work, we focus on FL, a communication-efficient and privacy-friendly approach to collaborative and distributed training of ML models.
Private distributed learning can also be built from transfer learning, as in~\cite{papernot2016semi,papernot2018scalable}.
The main intuition is to train a student model by transferring, through noisy aggregation, the knowledge of an ensemble of teachers trained on the disjoint subsets of training data.
Whereas Shokri and Shmatikov~\cite{shokri2015privacy} use differentially private gradient updates.

Work in~\cite{geyer2017differentially,mcmahan2017learning} present differentially private approaches to FL to add client-level protection by hiding participants' contributions during training. 
Whereas, in LDP, DP mechanisms are applied at the record level to hide the contribution of specific records in a participant's dataset.
An LDP-based FL approach is presented in~\cite{truex2020ldp} where participants can customize their privacy budget, while~\cite{pihur2018differentially} uses it for 
spam classification.
To the best of our knowledge, our research is the first to experiment with LDP and CDP against white-box membership inference attacks in FL and demonstrate that both can be used as viable defenses for backdoor attacks.

\descr{DP and poisoning attacks.} Prior work has also discussed the use of DP to provide robustness in ML, although not in FL as done in this paper. 
Ma et al.~\cite{ma2019data} show that DP can be effective when the adversary is only able to poison a small number of items, while Jagielski et al.~\cite{jagielski2020auditing} experiment with DP-SGD~\cite{abadi2016deep} against data poisoning attacks while assessing privacy guarantees it provides. 
Also, Hong et al.~\cite{hong2020effectiveness} introduce gradient shaping to bound gradient magnitudes, and experiment with DP-SGD in a non-FL setting, finding it to successfully defend against targeted poisoning attacks.
Overall, theese do not consider white-box inference attacks nor, more importantly, FL settings. 
Finally, Cheu et al.~\cite{cheu2021manipulation} explore manipulation attacks in LDP and evaluate lower bounds on the degree of manipulation allowed by local protocols for various tasks.

\section{Discussion \& Conclusion}
\label{discussion}
\reduce
Attacks against Federated Learning (FL) techniques have highlighted weaknesses in both robustness and privacy~\cite{kairouz2019advances}.
As for the former, we focused on backdoor attacks~\cite{bagdasaryan2020backdoor}; for the latter, on membership~\cite{nasr2019comprehensive} and property inference~\cite{melis2019exploiting} attacks.
To the best of our knowledge, prior work has only focused on protecting {\em either} robustness {\em or} privacy.
(Moreover, the latter has not experimented against white-box membership inference attacks such as the one presented in~\cite{nasr2019comprehensive}).

Aiming to provide both, our work was the first to investigate the use of Local and Central Differential Privacy (LDP/CDP) to mitigate both backdoor and inference attacks in FL.
Our intuition was that CDP limits the information learned about a specific participant, while LDP does so for records in a participant's dataset; in both cases, this limits the impact of poisonous data.
Overall, our work introduced the first analytical approach to empirically understand the effectiveness of LDP and CDP on protecting FL, also vis-\`a-vis the utility they provide in real-world tasks.

\descr{LDP.} Our experiments showed that LDP can successfully reduce the success of both backdoor and membership inference attacks.
For the former, LDP reduces attack accuracy further than state-of-the-art techniques such as clipping the norm of the gradient updates (``norm bounding'') and adding Gaussian noise (``weak DP'')~\cite{sun2019can}, although with a moderate cost in terms of utility.
\edit{For instance, as showed in Section~\ref{settingone}, LDP ($\epsilon=3$) for EMNIST with 2,400 participants, mitigates the backdoor accuracy from 88\% to 10\%, while the utility is reduced from 92\% to 62\%.}

\edit{In a more FL-suited dataset, i.e., Reddit-comments, with 51,548 participants, LDP ($\epsilon{=}1.7$) decreases  backdoor accuracy for task1, task2, and task3 from 83\%, 78\%, and 81\% to around 45\%, 43\%, and 55\%, with a reduction of utility from 19\% to 15\%.}
However, this only works against an adversary that is assumed to be able to modify her model updates but not the algorithm running on her device; to some extent, this is akin to a {\em semi-honest} adversary.
Whereas, if a {\em fully malicious} adversary does not add noise to her updates (i.e., she ``opts out'' from the LDP protocol), this could actually boost the accuracy of the backdoor attack.
Our experiments in Section~\ref{settingtwo} confirmed this was the case, as applying DP constrains the set of possible solutions in the optimization problem during training on the participant's dataset.
That is, not applying DP means the optimization problem has a larger space of solutions, and so any participant that does not apply DP can potentially have a bigger impact on the aggregated model.

As for privacy, LDP is effective against membership inference -- specifically, the white-box attack presented in~\cite{nasr2019comprehensive} -- reducing the adversary's accuracy without destroying utility.
For instance, LDP ($\epsilon=8.6$) reduces the accuracy of a global active attack from 91\% to 53\%, with utility going from 82\% to 68\% in CIFAR100.
However, LDP does not protect against property inference~\cite{melis2019exploiting}; this is not surprising since LDP only provides record-level privacy.

\descr{CDP.} CDP also provides a viable defense for backdoor and membership inference attacks.
\edit {In fact, CDP proved to be more effective than LDP against the former to reduce the backdoor accuracy better while providing a greater utility.
For instance, experiments in Section~\ref{settingone} showed that CDP ($\epsilon=3$) in EMNIST reduced the backdoor accuracy from 88\% to 6\%, which is better mitigation in comparison to LDP ($\epsilon=3$) that is 10\%. 
Besides, utility only reduced from 90\% to 78\% (higher than 62\% for LDP ($\epsilon=3$).}

As for privacy, in Section~\ref{subsec:defenses}, we found that CDP ($\epsilon=5.8$) reduces local active attack accuracy from 68\% to 55\%. However, utility goes down from 84\% to 70\%. 
In Texas100, CDP ($\epsilon=5.8$) mitigates the local passive attack by reducing the attack accuracy from 62\% to 54\%. 

Alas, we also found that CDP does not provide strong mitigations against property inference attacks in settings where the number of participants is small.
In other words, we can only obtain privacy {\em or} utility.
One might argue that FL applications like those deployed by Google~\cite{hard2018federated, yang2018applied} or Apple~\cite{ramaswamy2019federated} are likely to involve a number of participants in the order of thousands if not millions; however, it is becoming increasingly popular to advocate for FL approaches in much ``smaller'' applications, e.g., for medical settings~\cite{sheller2018multi,brisimi2018federated,jochems2}.

All in all, our experiments showed that we could obtain reasonable accuracy with LDP and CDP while reducing the performance of membership inference attack in FL.
However, remind that we cannot compare privacy bounds provided by LDP and CDP as they capture different concepts.

Overall, we are confident that our framework can be re-used to experiment with LDP and CDP along many more axes, such as different distributions of features and samples, complexity of the main tasks, number of participants, etc.

\descr{Limitations \& Future Work.} \ndss{In terms of robustness, we only look at backdoor attacks -- i.e., a {\em subset} of poisoning attacks.
Also, the evaluation of membership and property inference attacks follow, respectively,~\cite{nasr2019comprehensive} and~\cite{melis2019exploiting}, thus datasets and tasks we experiment with are limited to those from~\cite{nasr2019comprehensive,melis2019exploiting}. 
}

As part of future work, we plan to extend the robustness and privacy experiments to additional tasks and datasets and experiment with more attacks like model inversion and reconstruction attacks. 
\ndss{Overall, we are confident that our experimental framework can be extended to support combining multiple defenses addressing multiple robustness and privacy properties in FL, e.g., Byzantine-resilient aggregation algorithms for robustness with other techniques for inference and reconstruction attacks.}
Finally, we call for further work to provide more practical approaches to compare CDP and LDP; at the moment, this is not straightforward as these two DP variants entail different privacy properties, and theoretical techniques to convert one to the other, such as using group privacy, are not entirely practical.

\descr{Acknowledgments.} The authors wish to thank Boris K\"opf, Santiago Zanella-B\'eguelin, and Shruti Tople for helpful feedback and comments.
This work has been partially supported by a Microsoft EPSRC Case Studentship and an Amazon Research Award grant.

{\small
\bibliographystyle{abbrv}
\bibliography{ref}
}

\begin{table*}[t]
\begin{center}
\begin{tabular}{l|rr|rr|rr|rr|r|r|rr|rr}
\textbf{Hyperparameter} & \multicolumn{4}{c|}{\textbf{EMNIST}}  & \multicolumn{4}{c|}{\textbf{CIFAR10}} & \multicolumn{2}{c|}{\textbf{Reddit}}&\multicolumn{4}{c|}{\textbf{\ndss{Sentiment140}}} \\[0.5ex] %
\cline{2-15}
\rule{0pt}{1.05\normalbaselineskip} & \multicolumn{2}{c|}{\textbf{LDP}} & \multicolumn{2}{c|}{\textbf{CDP}} & \multicolumn{2}{c|}{\textbf{LDP}} & \multicolumn{2}{c|}{\textbf{CDP}} & \multicolumn{1}{c|}{\textbf{LDP}} & \multicolumn{1}{c|}{\textbf{CDP}} & \multicolumn{2}{c|}{\textbf{\ndss{LDP}}} & \multicolumn{2}{c|}{\textbf{\ndss{CDP}}}\\
\midrule
$\sigma$ (Noise Magnitude)	&0.8	&0.1	&-	&-	&0.5	&0.01 &- &- & 0.025&-&  \ndss{0.8}&\ndss{0.1}&\ndss{-}&\ndss{-} \\ 
S (Clipping Bound)	&5.0	 &5.0 	&3.0	 &5.0	&10.0 	&10.0 &10.0 &15.0 &5.0 &10.0   &\ndss{5}&\ndss{10}&\ndss{10}&\ndss{15}        \\ 
z (Noise Scale)	&- 	&-  &2.5 	&1.0 	&- 	&- &1.4 &1.0 &- &1.0  &\ndss{-}&\ndss{-}&\ndss{2.3}&\ndss{1.1}          \\ 
$\epsilon$ (Privacy Budget)	&3.0 	 &7.5 	&3.0 	&8.0 	&2.5 	&7.0 &2.8 &6.0 &1.7&1.2    &\ndss{1.9}&\ndss{6.7}&\ndss{2.2}&\ndss{6.1}        \\ 

\end{tabular}
\end{center}
\reduce
\caption{Hyperparameters for LDP and CDP in the Robustness Experiments.} 
\label{robustnessldpcdpparameters}
\end{table*}

\begin{table*}[t]
\setlength{\tabcolsep}{4pt}
\begin{center}
\begin{tabular}{l|r|r|r|r|r|r|rr}
\textbf{Hyperparameter} & \multicolumn{2}{c|}{\textbf{CIFAR100}}  & \multicolumn{2}{c|}{\textbf{Purchase100}} & \multicolumn{3}{c}{\textbf{LFW}}\\[0.5ex] %
\cline{2-8}
\rule{0pt}{1.05\normalbaselineskip}  & \multicolumn{1}{c|}{\textbf{LDP}} & \multicolumn{1}{c|}{\textbf{CDP}} & \multicolumn{1}{c|}{\textbf{LDP}} & \multicolumn{1}{c|}{\textbf{CDP}} & \multicolumn{1}{c|}{\textbf{LDP}} & \multicolumn{2}{c}{\textbf{CDP}}\\       
\midrule      
+$\sigma$ (Noise Magnitude)	&0.1	&-	&0.1	&-	&0.1	&- &-          \\ 
S (Clipping Bound)	&10.0	 &15.0 	&5.0	 &15.0	&12.0 	&10.0 &10.0            \\ 
z (Noise Scale)	&- 	&0.8  &- 	&1.1 	&- 	&1.4 &0.7            \\ 
$\epsilon$ (Privacy Budget)	&8.6 	 &5.8 	&8.6 	&5.8 	&10.7 	&4.7 &8.1            \\ 
\end{tabular}
\end{center}
\reduce
\caption{Hyperparameters for LDP and CDP in the Privacy Experiments.}
\label{privacyldpcdpparameters}
\end{table*}

\begin{table*}[t]
\centering
\begin{tabular}{r|r|r|r|r}
\textbf{\#Participants} & \textbf{No Defense} & \textbf{LDP} & \textbf{CDP} & \textbf{CDP} \\
&  & \textbf{($\epsilon$ = 10.7)} & \textbf{($\epsilon$ = 4.7)} & \textbf{($\epsilon$ = 8.1)} \\ \midrule

5                               & 90\%              & 83\%                      & 59\%                      & 85\%                       \\ 
10                              & 89\%              & 81\%                      & 57\%                      & 83\%                       \\ 
15                              & 88\%              & 80\%                      & 54\%                      & 82\%                       \\ 
20                              & 87\%              & 78\%                      & 53\%                      & 79\%                       \\ 
25                              & 85\%              & 70\%                      & 53\%                      & 77\%                       \\ 
30                              & 81\%              & 68\%                      & 51\%                      & 73\%                       \\ 
\end{tabular}
\caption{Main Task (Gender Classification) Accuracy with No Defense, LDP, and CDP (property inference attack setting).}
\label{lfwmodelaccuracy}
\end{table*}

\begin{table*}
\centering
\begin{tabular}{r|r|r|r}
\textbf{\#Participants} & \textbf{No Defense} & \textbf{LDP} & \textbf{CDP} \\ 
&  & \textbf{($\epsilon$ = 10.7)} & \textbf{($\epsilon$ = 8.1)} \\ \midrule

5                               & 0.97              & 0.95                      & 0.94                       \\ 
10                              & 0.87              & 0.86                      & 0.85                       \\ 
15                              & 0.76              & 0.75                      & 0.76                       \\
20                              & 0.70              & 0.70                      & 0.68                       \\ 
25                              & 0.54              & 0.52                      & 0.50                       \\ 
30                              & 0.48              & 0.47                      & 0.45                       \\ 
\end{tabular}
\reduce
\caption{AUC of the Property Inference Attack in~\cite{melis2019exploiting} with No Defense, LDP, and CDP.}
\label{lfwattackaccuracy}
\end{table*}

\appendix

\section{Hyperparameters}\label{sec:hyper}

In Tables~\ref{robustnessldpcdpparameters} and~\ref{privacyldpcdpparameters}, we report hyperparameters for LPD and CDP in, respectively, the robustness and privacy experiments.

\section{Property Inference Attack}\label{sec:propertyinferenceattack}
Table~\ref{lfwmodelaccuracy} presents the federated gender classification main task accuracy when LDP, CDP, and no DP are applied. 
Table~\ref{lfwattackaccuracy} reports the property inference attack accuracy in the form of the area under a curve (AUC score) when LDP, CDP, and no DP are applied.

\end{document}